\documentclass[AMA,LATO1COL]{WileyNJD-v2}
\usepackage{moreverb}

\newcommand\BibTeX{{\rmfamily B\kern-.05em \textsc{i\kern-.025em b}\kern-.08em
T\kern-.1667em\lower.7ex\hbox{E}\kern-.125emX}}

\articletype{Article Type}%

\received{<day> <Month>, <year>}
\revised{<day> <Month>, <year>}
\accepted{<day> <Month>, <year>}

\usepackage{mathrsfs,subfigure}
\usepackage{array}
\usepackage{subfigure}
\usepackage{xcolor}
\usepackage{hyperref}
\usepackage{url}
\usepackage{booktabs,tabularx,colortbl,threeparttable}
\usepackage{multirow,amsmath,mathrsfs,verbatim,bbm}
\usepackage{graphicx}
\definecolor{mygray}{gray}{.7}
\usepackage{listings}
\usepackage{xspace}
\usepackage{adjustbox}
\usepackage{colortbl}

\usepackage{amssymb}

\begin{document}

\title{Characterizing Commodity Serverless Computing Platforms}

\author[1,2]{Jinfeng Wen}

\author[1]{Yi Liu*}

\author[3]{Zhenpeng Chen}

\author[1]{Junkai Chen}

\author[2]{Yun Ma}


\authormark{Wen \textsc{et al.}}

\address[1]{\orgdiv{Key Lab of High-Confidence Software Technology, MoE}, \orgname{Peking University}, \orgaddress{\state{Beijing}, \country{China}}}

\address[2]{\orgdiv{Institute for Artificial Intelligence}, \orgname{Peking University}, \orgaddress{\state{Beijing}, \country{China}}}

\address[3]{\orgdiv{Department of Computer Science}, \orgname{University College London}, \orgaddress{\state{London}, \country{UK}}}



\corres{*Yi Liu, Key Lab of High-Confidence Software Technology, MoE, Peking University, Beijing, China. \email{liuyi14@pku.edu.cn}}










\newcommand{\nickName}{\textit{SlsBench}\xspace}

\newcommand{\testbedName}{\textit{TBS}\xspace}

\newcommand{\para}[1]{\smallskip\noindent{\bf {#1}. }}

\newcommand{\wjf}[1]{{\color{red}{#1}}}
\newcommand{\ly}[1]{{\color{blue}{#1}}}

\newcommand{\tabincell}[2]{\begin{tabular}{@{}#1@{}}#2\end{tabular}} 


\abstract[Abstract]{Serverless computing has become a new trending paradigm in cloud computing, allowing developers to focus on the development of core application logic and rapidly construct the prototype via the composition of independent functions. With the development and prosperity of serverless computing, major cloud vendors have successively rolled out their commodity serverless computing platforms. However, the characteristics of these platforms have not been systematically studied. Measuring these characteristics can help developers to select the most adequate serverless computing platform and develop their serverless-based applications in the right way. To fill this knowledge gap, we present a comprehensive study on characterizing mainstream commodity serverless computing platforms, including AWS Lambda, Google Cloud Functions, Azure Functions, and Alibaba Cloud Function Compute. Specifically, we conduct both qualitative analysis and quantitative analysis. In qualitative analysis, we compare these platforms from three aspects (i.e., development, deployment, and runtime) based on their official documentation to construct a taxonomy of characteristics. In quantitative analysis, we analyze the runtime performance of these platforms from multiple dimensions with well-designed benchmarks. First, we analyze three key factors that can influence the startup latency of serverless-based applications. Second, we compare the resource efficiency of different platforms with 16 representative benchmarks. Finally, we measure their performance difference when dealing with different concurrent requests, and explore the potential causes in a black-box fashion. Based on the results of both qualitative and quantitative analysis, we derive a series of findings and provide insightful implications for both developers and cloud vendors.




}



\maketitle



\section{Introduction}\label{intro}

Serverless computing is a new paradigm of cloud computing that is gaining traction in a wide range of domains including video processing\cite{ao2018sprocket}, machine learning\cite{carreira2019case}, scientific computing\cite{shankar2020numpywren}, etc. It is predicted that 50\% of global enterprises will employ serverless computing by 2025\cite{gartner20new}. With serverless computing, developers can focus on only the application logic composed by dependent functions, which are small pieces of the program dedicated to a simple task, i.e., Function-as-a-Service (\emph{FaaS})\cite{JonasCoRR2019}. Therefore, developers can be freed from tedious and error-prone infrastructure management such as load-balancing and auto-scaling. Meanwhile, developers can also directly integrate existing proprietary services, i.e., Backend-as-a-Service (\emph{BaaS})\cite{JonasCoRR2019}, provided by cloud vendors like object storage service (Amazon S3)\cite{AWSS3}. This new paradigm can reduce the cost as developers pay for only the actual function executions, and bring great benefits for cloud vendors as it allows them to better utilize resources. Currently, major cloud vendors have rolled out their serverless computing platforms such as AWS Lambda\cite{awsnew}, Google Cloud Functions\cite{googlenew}, Azure Functions\cite{azurenew}, and Alibaba Cloud Function Compute\cite{alibabanew}.


These commodity serverless computing platforms often act in a black-box fashion, and developers do not need to pay attention to the underlying implementation details. As a result, it may be difficult for developers to select the most appropriate serverless computing platform to host their applications or construct their applications in the right way on serverless computing platforms. For example, how can developers evaluate whether or how their existing applications can be transformed into serverless-based functions\cite{frey2013automatic}?  How can developers package their newly developed functions on a specific platform? How many configuration options does a serverless computing platform provide for developers to quickly meet their requirements\cite{Wen2021ServerlessWorkflow}? How about the actual runtime performance of these serverless computing platforms for specific tasks? These issues are non-trivial in practice as they can definitely impact the developer's decision-making, and thus potentially impact the quality of service, user experience, and even the revenue of the application.

Unfortunately, to the best of our knowledge, there still lacks the comprehensive knowledge to assist developers to select the most appropriate serverless computing platform and develop their serverless-based applications in the right way. In this paper, we conduct a comprehensive study with both qualitative analysis and quantitative analysis to characterize four mainstream commodity serverless computing platforms, including AWS Lambda, Google Cloud Functions, Azure Functions, and Alibaba Cloud Function Compute. 

In qualitative analysis, we explore and summarize the various characteristics described in the official documentation of these serverless computing platforms. These characteristics specify the inherent restrictions of different aspects involving developing, deploying, and executing functions, which may result in fatal failures if developers do not comply with these restrictions. To this end, we construct a taxonomy with respect to the need-to-consider information from three aspects,i.e., development, deployment, and runtime. Such a taxonomy can help developers better understand the supported characteristics of serverless computing platforms to facilitate further development practice.

In quantitative analysis, we explore the actual runtime performance of these serverless computing platforms from multiple dimensions, in order to help developers select an appropriate platform based on their application features and improve applications' performance with tuned configurations. Indeed, the overall perceived performance of a serverless application may mainly be influenced by three kinds of latency, i.e., the startup latency of initiating the function instance, the execution latency of running the function, and the scheduling latency of waiting for serving by available instances when the number of requests dramatically increases. Thus, our approaches are designed as follows.

\begin{itemize}
\item First, we quantitatively analyze how can programming languages, memory sizes, and package sizes influence the startup latency on different serverless computing platforms. Startup latency severely affects the responsiveness of serverless applications and may limit the adoption of serverless computing under various applications.

\item Second, we quantitatively measure the applications' actual runtime performance to compare the underlying resource efficiency of different serverless computing platforms with a set of well-designed benchmarks. We categorize these benchmarks into two types,i.e., microbenchmarks and macrobenchmarks. Microbenchmarks consist of a set of simple workloads focusing on specific resource consumption, such as CPU, memory, network, disk IO, etc. Macrobenchmarks consist of a set of real-world representative applications, e.g., multimedia data process, MapReduce, machine-learning-based serving, which need to utilize various system resources.

\item Finally, we quantitatively compare the concurrency performance of different serverless computing platforms, i.e., how they perform when dealing with multiple requests due to different auto-scaling features and inherent concurrency limits. A coming request may be throttled if no available function instances can handle it, which results in non-negligible latency. Meanwhile, we try to analyze the potential causes influencing concurrency performance by analyzing and inferring their scalability strategy and load balancing from a black-box perspective. 

\end{itemize}

Based on the results of both qualitative and quantitative analysis, we report a series of findings and implications for developers and cloud vendors. Our findings can not only help developers choose the right platforms and configurations to obtain the optimal performance based on their actual workloads, but also guide cloud vendors to improve their serverless computing platforms.

We divide the process of using serverless computing into three phases, i.e., development, deployment, and runtime. Our main findings include: 

\noindent $\bullet$ \textbf{Development.} Programming languages are yet limitedly supported on most serverless computing platforms. Only nine popular programming languages along with a small portion of versions are supported on at least one serverless computing platform, and three (PowerShell, PHP, and TypeScript) out of nine languages are supported on only one platform. Especially, developers need to consider the supported programming language before transforming a legacy application (a.k.a., serverful application) to a serverless application. Additionally, different languages can result in obviously different cold start latency. Languages with ``cumbersome'' runtime (e.g., Java) can account for a much longer cold start time (7x) than other script languages (e.g., Python and Node.js). Due to the package size limit of serverless computing platforms, developers need to control their package size when constructing their applications. For example, Google Cloud Functions does not support applications with more than 500 MB uncompressed package size, which means that developers even cannot directly use some most popular but heavy libraries like ``TensorFlow 2.4''.

\noindent $\bullet$ \textbf{Deployment.} When deploying serverless-based functions, we find that loading redundant libraries can introduce non-negligible startup latency as well as increase package size. The startup latency can even increase by 4.6x if the function loads some unused libraries. Therefore, developers should optimize their code (e.g., removing the useless code) and apply the lazy loading of libraries to alleviate the overhead of initializing functions. It is better for cloud vendors to provide related tools to help developers optimize their function packages before development instead of directly deploying the compressed function packages. In addition, each serverless computing platform has different limitations of configuration options, such as function memory and timeout. For example, AWS Lambda has currently supported up to 10,240 MB memory and 900 seconds function timeout. We also need to address that existing serverless computing platforms do not support GPU-enabled applications. Developers can carefully consider whether their applications can successfully run with limited system resources and permitted configurations on the target serverless computing platform.

\noindent $\bullet$ \textbf{Runtime.} Allocating more memory within a certain range can obviously reduce the cold start time. For example, increasing the memory from 128 MB to 2,048 MB can save more than 98.5\% of the cold start time for a task running on Google Cloud Functions. For tasks with high memory requirements (e.g., \textit{sls-fib}), allocating more memory can also reduce the execution time dramatically. However, the rate of reduction will drop when allocated memory exceeds 1,024 MB. These findings help developers make a trade-off between cost and performance by allocating memory adequately. We also find that different serverless computing platforms can have discrepant performance for a specific type of task. For example, Alibaba Cloud Function Compute performs the best for CPU-bound and memory-bound benchmarks, but other platforms can beat it in other benchmarks with certain conditions, such as Google Cloud Functions has higher random-IO throughput at 1,024 MB of memory. In addition, for concurrency tasks, AWS Lambda has stronger ability of concurrency requests than other platforms. However, the concurrency performance of Alibaba Cloud Function Compute is the best and most stable. For example, compared with the concurrency performance of AWS Lambda, Google Cloud Functions, and Azure Functions, Alibaba Cloud Function Compute can improve by from 4.29\% to 102.50\%, from 37.27\% to 151.16\%, and from 271.31\% to 445.82\%, respectively. Through analyzing the difference in concurrency performance, we find that the potential causes may be the different scalability strategies of these platforms. We also find that the memory affects the concurrency performance of Alibaba Cloud Function Compute. The larger the memory, the better the concurrency performance. For example, the performance of concurrency tasks that each request allocates 128 MB is inferior to that of allocating 3,072 MB by about from 60\% to 80\%.

Overall, our findings reveal the mystery of mainstream commodity serverless computing platforms, motivating future research and application practice of serverless computing. We also provide insightful implications for developers and cloud vendors. 
In addition, we offer the experimental code used in this study\footnote{\url{https://github.com/WenJinfeng/TBS}} as an additional contribution to the research community for other researchers to replicate and build upon.



The remainder of this paper is organized as follows. Section~\ref{sec:methodlogy} presents our methodology. Section~\ref{sec:char} summarizes and compares the key characteristics of development, deployment, and runtime phases. Section~\ref{sec:testbed} describes the details of our evaluation tool. Section~\ref{sec:evaluation} shows the evaluation results of the actual runtime performance for four serverless computing platforms. Section~\ref{sec:discussion} discusses the potential issues of our study. Section~\ref{sec:related_work} surveys
related work and Section~\ref{sec:conclusion} concludes this work.



\section{Methodology}\label{sec:methodlogy}


This section illustrates the methodology that we adopt to evaluate mainstream commodity serverless computing platforms, including Amazon Web Service Lambda (released in November 2014), Google Cloud Functions (released in July 2018), Microsoft Azure Functions\footnote{Azure Functions offers three different hosting plans~\cite{azurenew}, we focus on only the consumption plan that is the most similar to other serverless computing platforms.} (released in November 2016), and Alibaba Cloud Function Compute (released in April 2017). 
To help developers comprehensively understand these serverless computing platforms, we compare them via both qualitative analysis and quantitative analysis.

In qualitative analysis, we aim to help developers to gain an intuitive understanding of serverless computing and judge whether their applications can be implemented and deployed on a specific serverless computing platform. 
In quantitative analysis, we seek to explore the actual runtime performance of these serverless computing platforms from multiple dimensions. Indeed, the overall perceived performance of serverless applications is mainly influenced by the startup latency (initiating the function instance), the execution latency (running the function), and scheduling latency (waiting for serving by available instances when the number of requests dramatically increases). Specifically, first, functions are typically small and executed in seconds or even milliseconds, thereby the startup latency can be a considerable overhead for overall performance. Moreover, executing serverless applications with low startup latency is critical for user experience\cite{HellersteinCIDR19, DuASPLOS2020, WenServerless21, lahmar2018multicloud}. Second, different kinds of tasks have various resource demands, and it is vital for developers to understand the execution performance of these tasks to choose the most appropriate platform. Finally, the scheduling strategy of different serverless computing platforms can affect the response time from the time of receiving a new request to the time of allocating a dedicated instance. Moreover, the main advertised benefit of serverless computing is the automatic scaling, we wonder if they can perform as stated. Therefore, based on the above analysis, we mainly explore the actual runtime performance of these platforms from three dimensions, including startup latency, execution latency, and scheduling latency.


\begin{figure}[htbp]
\centering
\begin{minipage}[t]{0.48\textwidth}
\centering
\includegraphics[width=0.98\textwidth]{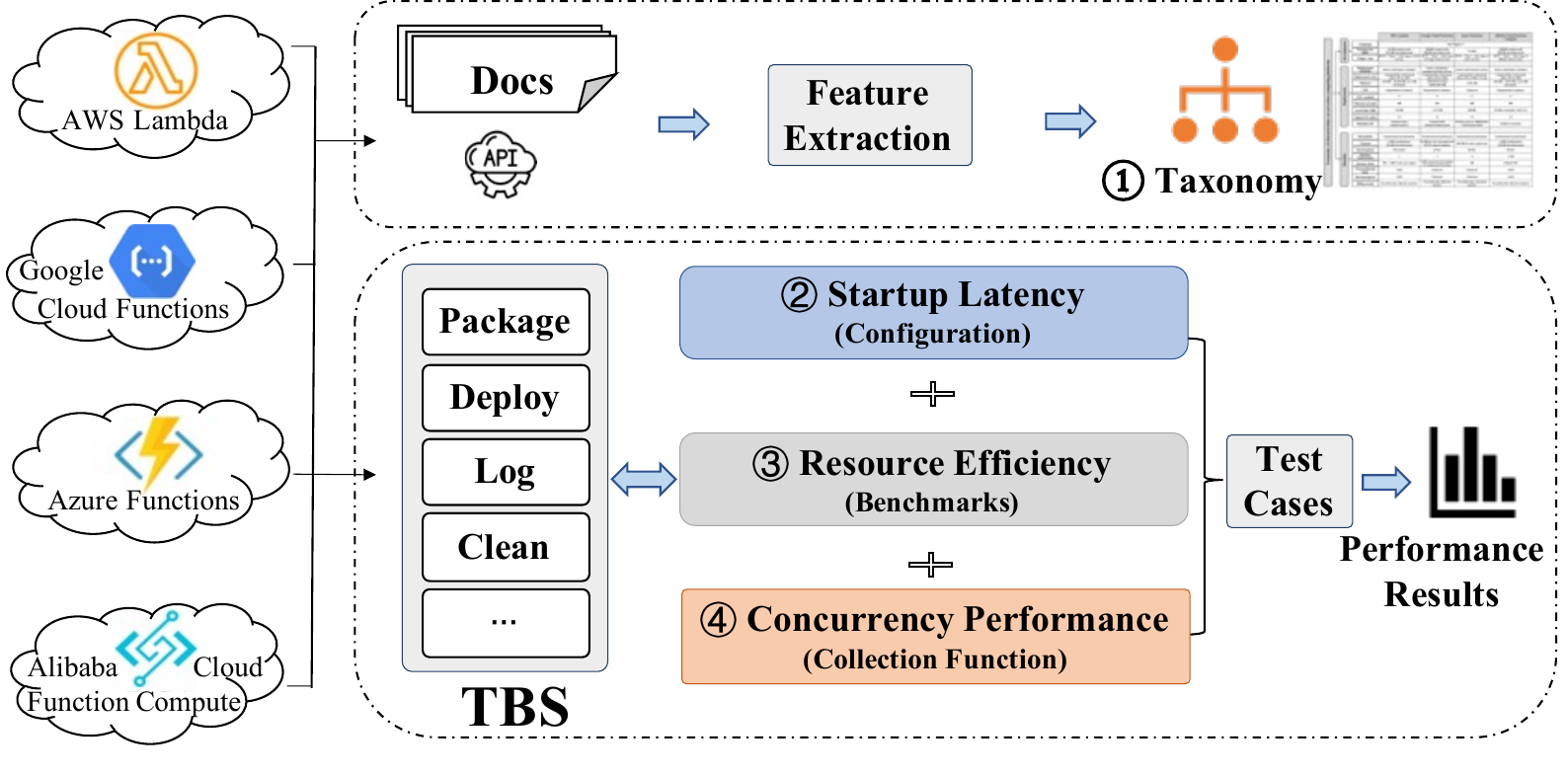}
\caption{An overview of the methodology.}
\label{fig:workflow}
\end{minipage}
\begin{minipage}[t]{0.48\textwidth}
\centering
\includegraphics[width=0.98\textwidth]{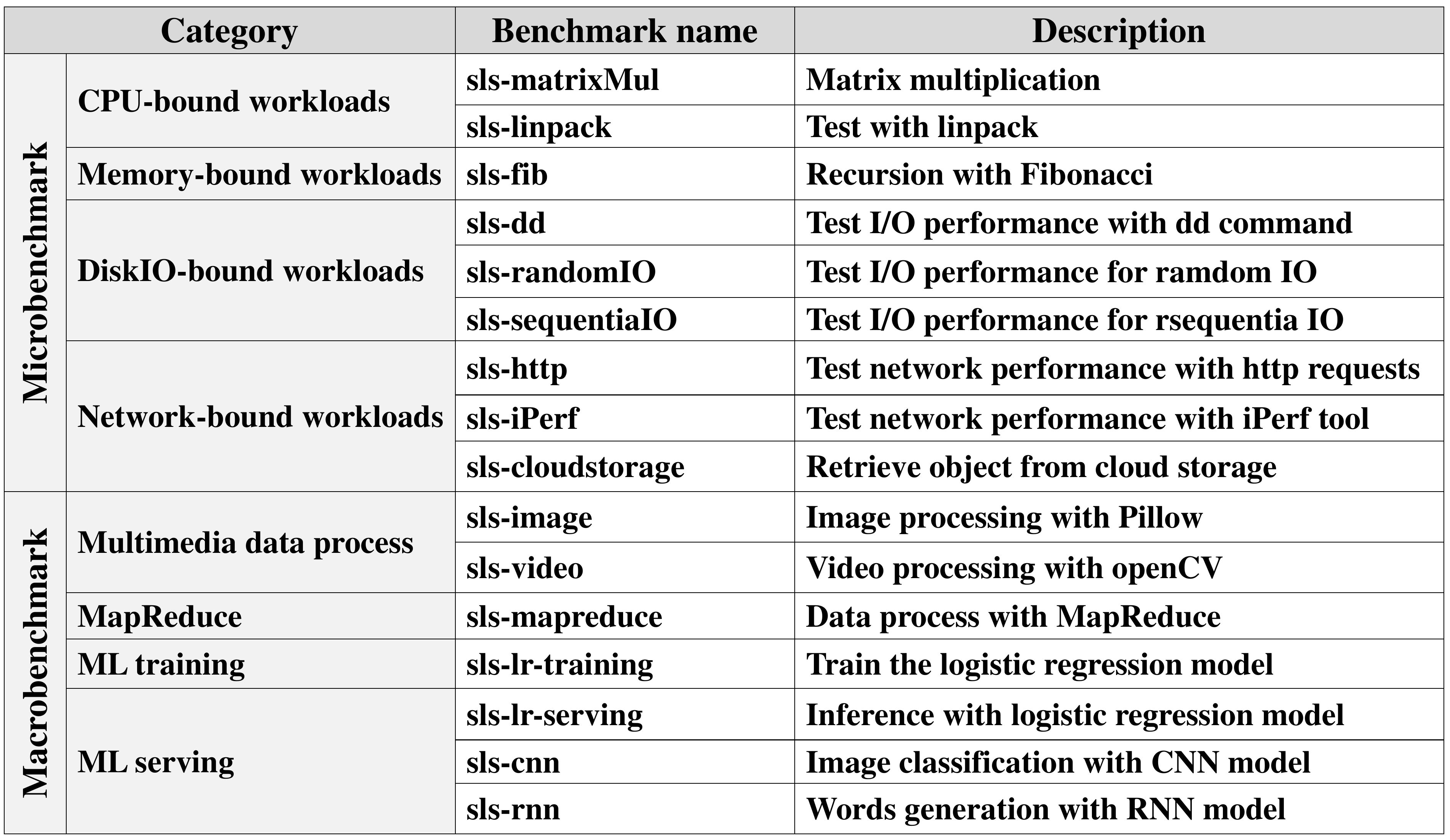}
\caption{Benchmarks.}
\label{fig:benchmark}
\end{minipage}
\end{figure}




Figure~\ref{fig:workflow} shows the overall of our methodology. Step \textcircled{1} is the qualitative analysis, which is to extract and summarize characteristics of four serverless computing platforms from their official documentation. Such characteristics specify the inherent restrictions of different aspects, e.g., development, deployment, and runtime, and a taxonomy of characteristics about these platforms is constructed. Detailed results are shown in Section \ref{sec:char}. In our methodology, steps \textcircled{2} \textcircled{3} \textcircled{4} are the quantitative analysis, which explores the actual runtime performance on these serverless computing platforms from multiple dimensions. Before experimental evaluation, we first design and implement an evaluation tool named \textbf{\testbedName}, which uses a design principle of modularity and extensibility to allow developers to integrate other platforms without restraint. \testbedName can provide the function packaging, deployment, execution, log collection, clean-up, and result generation for each measurement on four serverless computing platforms. Based on \testbedName, we conduct our evaluation and first explore the key factors (e.g., programming languages, memory sizes, and package sizes) influencing startup latency through varied experimental configurations (step \textcircled{2}). Detailed results are shown in Section \ref{sec:startuplatency}. Second, we construct a benchmark suite including different types of benchmarks to investigate the performance of various tasks with different resource demands on these platforms (step \textcircled{3}). Detailed results are shown in Section \ref{sec:resourceefficiency}. Finally, we measure the concurrency performance of different numbers of concurrent requests on serverless computing platforms (step \textcircled{4}). Moreover, we construct a runtime information collection function to obtain some underlying information of platforms to further analyze potential causes influencing concurrency performance. The collection function is a serverless function to collect information like execution timestamp, function instances, virtual machines (VMs), etc. The collected information can reflect the scalability strategy and load balancing ability of platforms to a certain extent. Detailed results are shown in Section \ref{sec:scalability}.

\section{Taxonomy of Characteristics}\label{sec:char}

\begin{figure*}[!thb]
	\centering
    \includegraphics[width=0.9\textwidth]{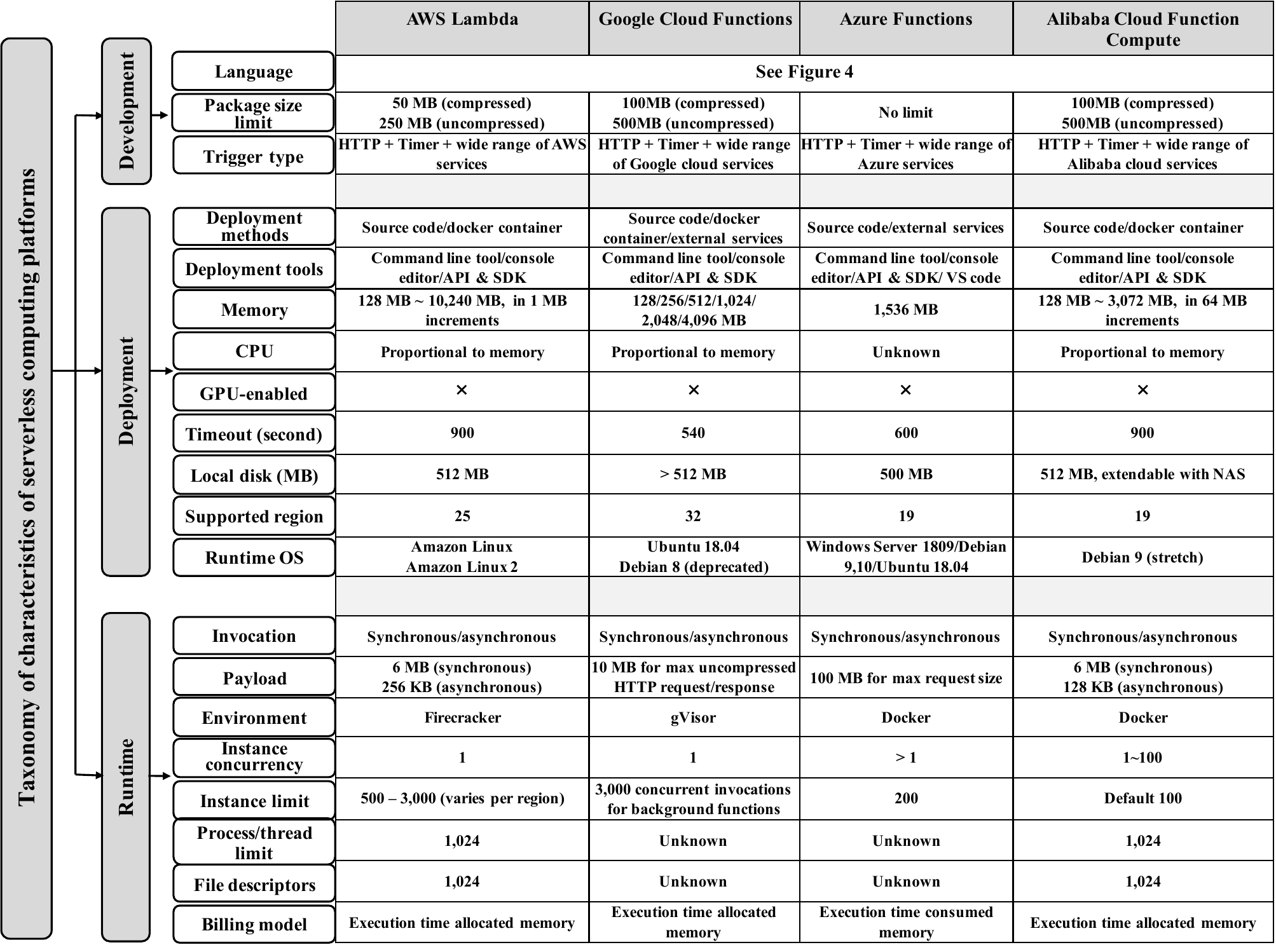}
    \caption{Taxonomy of characteristics of serverless computing platforms.}
    \label{fig:info}
\end{figure*}

To intuitively understand the characteristics of serverless computing, we construct a taxonomy of characteristics related to the development, deployment, and runtime for four serverless computing platforms. This taxonomy is shown in Figure~\ref{fig:info}, and specific details are illustrated as follows.

\begin{figure*}[!thb]
	\centering
    \includegraphics[width=0.93\textwidth]{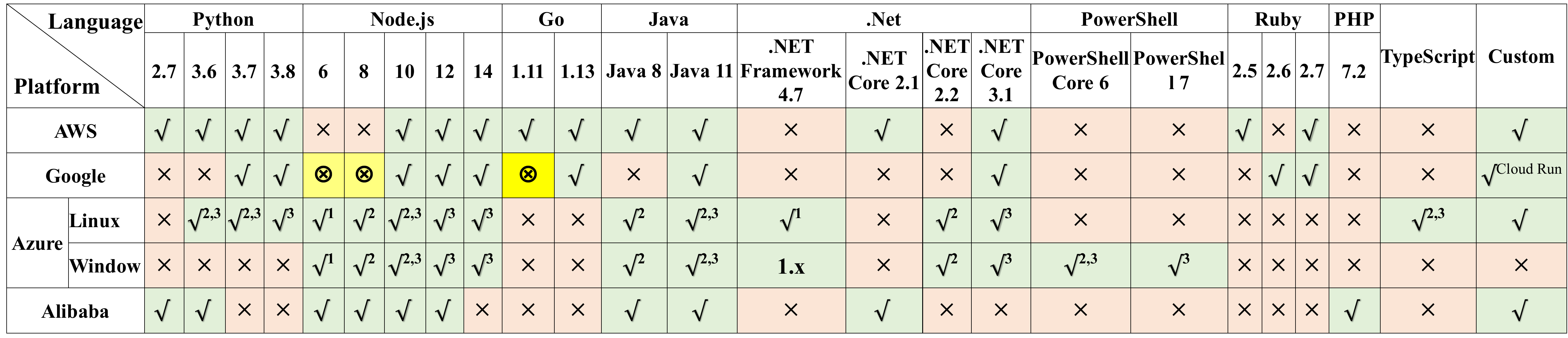}
    \caption{Supported languages of serverless computing platforms ($\checkmark$: supported; $\times$: unsupported; $^{1,2,3}$: generation of Azure Functions; \textcircled{$\times$}: deprecated.)}
    \label{fig:language}
\end{figure*}

\subsection{Development-related Characteristics}

When developing applications, developers need to consider whether a specific serverless computing platform can implement application functionalities in terms of programming language, trigger type, the size limit of deployment package, payload size, etc. The development-related characteristics are illustrated as follows.

Generally, a serverless-based application consists of one or more serverless functions. Each serverless function is a small, stateless, event-driven, and pay-as-you-go unit dedicated to handling a specific task. Such a function is often constructed with a small piece of code written in different languages. As shown in Figure~\ref{fig:language}, different serverless computing platforms support various \textbf{languages and versions}. We find that the three most popular languages (i.e., Python, Node.js, and Java\cite{toplanguagenew}) are natively supported by these serverless computing platforms. However, Python is not supported by Azure on Windows .Net Core. We also find that \emph{three languages (PowerShell, TypeScript, and PHP) are only natively supported by only one platform}. Specifically, PowerShell and TypeScript are only supported by Azure Functions, and PHP is only supported by Alibaba Cloud Function Compute. Since each platform supports various languages with different versions, we will choose the consistent language and version in our experimental evaluation, in order to compare their performance features fairly. In addition, four serverless computing platforms have the custom manner to allow developers to build their runtime with any preferred language, whereas may bring extra non-negligible efforts and concerns of runtime stability. Thus, it is important to choose the appropriate platform to develop applications based on developers' preferred languages.


In the specific implementation of application functionalities, these applications are usually triggered to execute by events\cite{JonasCoRR2019}, such as HTTP requests, timers, storage conditions. Besides, functions with different \textbf{event triggers} may have different main methods with given parameters, and they are invoked to deal with specific incoming events. To reduce the cold start time of serverless functions, serverless computing platforms often have the \textbf{package size limit} of applications. At the time of writing of our study, the package size limit of AWS Lambda is 50 MB with a compressed format and 250 MB with an uncompressed format. Google Cloud Functions and Alibaba Cloud Function Computing both allow from 100 MB with a compressed format to 500 MB with an uncompressed format. Differently, Azure Functions supports the higher package size as much as several GB in size. Unfortunately, the package size of applications, especially, deep-learning-based tasks with huge libraries, is big enough to exceed the package size limit of serverless computing platforms. For example, the compressed size of the newest ``TensorFlow'' library (version 2.4.1-9~\cite{tensorflownew}) is 106.4 MB, and the uncompressed installed size is as high as 688.1 MB. It is impossible to deploy such an application with the ``big'' library to serverless computing platforms directly, thus the development enthusiasm of developers may be dampened.

\noindent\textbf{Discussion and implications.} Developers should carefully select the appropriate serverless computing platform based on their preferred languages, the application package size, and other features that may be restricted on some platforms. For the application package limit, developers can apply existing technologies and tools\cite{RomanoTSE20} to remove the useless code and compress them. Cloud vendors can also provide some tools to automatically reduce the package size to ease the burden of developers, instead of directly deploying zipped packages submitted by developers.

\subsection{Deployment-related Characteristics}



After implementing serverless-based applications, developers need to deploy them to serverless computing platforms. During the deployment phase, some essential characteristics including deployment method and tool, allocated memory, function timeout, etc., need to be considered by the developers to avoid unnecessary deployment failure and poor performance. Next, we introduce this part of the information in detail.

As shown in Figure~\ref{fig:info}, we summarize the three most popular \textbf{deployment methods} of serverless applications, i.e., source code, docker container, and external services. On one hand, four serverless computing platforms all support the deployment method with the source code, e.g., deployment with a zipped package that contains function code and dependent third-party libraries. On the other hand, we find that only Azure Functions (with consumption plan) does not allow developers to deploy functions by building their custom runtime (e.g., building custom Docker images), but other platforms can. Additionally, Azure Functions and Google Cloud Functions both support the deployment with other external services, such as source control services (e.g., Git), FTP services, etc., whereas AWS Lambda and Alibaba Cloud Function Compute do not. When deploying functions to serverless computing platforms, developers can use \textbf{deployment tools} like the command-line tool, console editor, and API $\&$ SDK. Moreover, developers of Azure Functions can even deploy functions from specific development tools, e.g., Visual Studio Code.


In the deployment process, some serverless computing platforms (e.g., AWS Lambda, Google Cloud Functions, and Alibaba Cloud Function Compute) need to specify the \textbf{memory} size of functions in advance. Specifically, AWS Lambda allocates memory to a function instance from 128 MB to 10,240 MB in steps of 1 MB, Google Cloud Functions is from 128 MB to 4,096 MB with assigned values, and Alibaba Cloud Function Compute is from 128 MB to 3,072 MB in steps of  64 MB. However, for Azure Functions, developers cannot specify the memory in the development or deployment, but they can use 1,536 MB at most. Generally, memory and CPU are closely related. The \textbf{CPU} capability of a function instance increases proportionally with the allocated memory. AWS Lambda and Alibaba Cloud Function Compute get one vCPU at 1,769 MB and 1,024 MB, respectively. Regrettably, we find that \textbf{GPU support} for the function execution is not available in four serverless computing platforms. An alternative way is to provide GPU power as serverless services\cite{Jun18PDPnew, alibabaGPUnew}.

Each serverless computing platform allows a function to run within the \textbf{timeout} limit as shown in Figure~\ref{fig:info}. Thus, \textit{it is not recommended to run long-time tasks with serverless functions}. Specifically, the timeout limit of Google Cloud Functions is only 540 seconds, whereas both AWS Lambda and Alibaba Cloud Function Compute specify the maximum timeout time as long as 900 seconds. For Azure Functions, it allows functions to run within 600 seconds. In addition, serverless computing platforms also have the \textbf{local disk} to restrict the local storage (roughly 0.5 GB by default). However, Alibaba Cloud Function Compute enables developers to extend the local storage capacity with network-attached storage (NAS).

We collect the detailed information of the \textbf{supported regions} through deploying applications on each serverless computing platform factually. At the time of writing, AWS Lambda spans 25 geographical regions, whereas Google Cloud Functions supports the function deployment in 32 regions. Azure Functions and Alibaba Cloud Function Compute are available in 19 regions. However, \textit{we find an inconsistency compared to the official documentation of Azure Functions\footnote{Products available by region. \url{https://azure.microsoft.com/en-us/global-infrastructure/services/?products=functions&regions=all}},} which mentions that 43 regions overall are available. For the \textbf{runtime OS} of serverless computing platforms, most of them run applications on Linux hosts, and only Azure Functions supports both Windows hosts and Linux hosts. As shown in Figure~\ref{fig:language}, Windows hosts and Linux hosts do not support the same language set. In addition, we find that \textit{certain inconsistencies between the official documentation and the actual execution process of our experiments.} For example, AWS Lambda claims the usage of \textit{Amazon Linux} as the operating system for Python 3.7 runtime\footnote{\url{https://docs.aws.amazon.com/lambda/latest/dg/lambda-runtimes.html}}. However, it actually employs the \textit{Amazon Linux 2} as the operating system when we deploy applications with Python 3.7 runtime. Thus, we infer that the reason may be the belated update of AWS Lambda documentation, and this kind of inconsistency makes developers confused.




\noindent{\textbf{Discussion and implications.}} Since different serverless computing platforms have different memory capabilities, developers can choose an appropriate memory size to balance performance and cost. Meanwhile, with the increase of GPU requirements (e.g., deep learning tasks), cloud vendors are advised to timely solve the GPU technical difficulty and provide users with GPU as a service. Due to the timeout limit of functions, long-time applications are not suitable to run on serverless computing platforms. Developers can decompose their applications into multiple functions through runtime inspection. For the limit of local disk, data-intensive applications may also be limited. Cloud vendors are advised to improve the local storage size and provide a rapid cache mechanism to serve the data transformation. According to supported regions for serverless computing platforms, developers can choose the appropriate region near them to reduce the network latency.

\subsection{Runtime-related Characteristics}

Functions that have been successfully deployed on serverless computing platforms will be triggered and invoked when receiving defined events. Some runtime-related characteristics (e.g., invocation type, request payload, concurrency, etc) have to be taken into consideration by developers.

Specifically, four serverless computing platforms all support synchronous and asynchronous \textbf{invocations}. Moreover, developers often consider the request \textbf{payload} when calling functions, but the payload limit of each serverless computing platform is different and shown in Figure~\ref{fig:info}. When receiving bursty concurrent requests, serverless computing platforms can scale out by creating multiple function instances. A function instance often runs on a separate host \textbf{environment}, such as a lightweight VM (e.g., Firecracker\cite{AgacheNSDI2020} used by AWS Lambda), a dedicated container (e.g., Docker\cite{alibabanew} used by Alibaba Cloud Function Compute and Azure Functions), or a gVisor\cite{gvisornew} used by Google Cloud Functions. Furthermore, all the resources (e.g., CPU, memory, and storage) of this host are dedicated solely to its function instance. For \textbf{instance concurrency}, serverless computing platforms will launch multiple function instances to handle concurrent request tasks. Generally, each new request will launch one exclusive instance on AWS Lambda and Google Cloud Functions. However, a single instance on Azure Functions and Alibaba Cloud Function Compute (with the custom setting of concurrency number) can run handle multiple concurrent requests (i.e., starting multiple processes) at the same time. We will further explore these features in our later evaluation. Depending on the deployment region, the \textbf{instance limit} supported by AWS Lambda is from 500 up to 3,000 per single function. Google Cloud Functions only allows up to 3,000 invocations to be executed concurrently for background functions. For Azure Functions and Alibaba Cloud Function Computing, their instance limits are 200 and 100, respectively.

Serverless computing platforms allow developers to normally use language and operating system features, such as creating additional threads and processes. Resources allocated to serverless applications, including memory, disk, and network, must be shared among all the \textbf{processes/threads}. In addition, the Linux kernel uses \textbf{file descriptors} to efficiently manage the opened files and creates indexes for them. For the \textbf{billing model} of serverless computing platforms, the cost generally rises with the increase of execution time and the allocated memory. In serverless computing, developers can pay for their applications in a finer granularity, i.e., milliseconds. Indeed, the overall cost of running an application varies depending on the specified memory (in AWS Lambda, Google Cloud Functions, and Alibaba Cloud Function Compute) or the actually consumed memory during executions (in Azure Functions).

\noindent{\textbf{Discussion and implications.}} Developers should notice the payload limit when passing input data to applications. If the payload data is large, developers can leverage the external storage service (e.g., AWS S3) to storage it, and then trigger functions contained in applications. Due to the black-box feature of commodity serverless computing platforms, some runtime information and performance are not shown in their official documentation. It will drive us to further explore these runtime features including the actual performance and the key factors influencing performance in our later evaluation, in order to help developers better construct and deploy their functions.

\section{The evaluation tool}\label{sec:testbed}

To explore the actual runtime performance on four serverless computing platforms, we first design and implement an open-source and modular evaluation tool named \textbf{\testbedName}, which contains some necessary components and a benchmark suite with 16 representative benchmarks.

\testbedName abstracts necessary components for each serverless computing platform, including the packaging component, the deploying component, the testing component, the logging component, and the result generating component. The packaging component automatically packages a benchmark as a zipped file for further deployment. The deploying component can specify configurations (shown in Figure~\ref{fig:info}) and deploy functions to the target platform. The testing component is to execute deployed functions based on configured trigger types, e.g., sending an HTTP request. The logging component is used to retrieve and store the execution logs of each measurement. Although different serverless computing platforms have different formats of logs, they often contain basic information (e.g., timestamps, the execution time of functions, the output result of functions, etc.) about the life cycle of a function. For each benchmark in our study, we save the important contents, as well as the output result in logs. Result generating component can extract valuable information from structured logs, and generate a comparable result about different serverless computing platforms.
If developers want to integrate a new serverless computing platform into \testbedName, they just need to provide the platform-specific implementation with the official command-line tool, APIs, or SDK. Leveraging these basic components, developers can easily evaluate different platforms with specific configurations and experiment settings.

\testbedName also provides a benchmark suite containing different benchmarks. Each benchmark is implemented in multiple programming languages (Python by default) to support the variety of the available runtime systems. Although there are subtle differences between code variants to adapt to the peculiarities of each serverless computing platform, the code of each programming language is basically the same for all. By default, our functions are implemented using the HTTP trigger, which is the most common event trigger supported by all serverless computing platforms. The benchmark suite is considered to measure the runtime performance of these serverless computing platforms with workloads that have various system resource requirements, such as CPU, memory, disk IO, and network. In addition, real-world workloads that developers develop often consume extensive system resources to complete complex tasks. Therefore, we mainly design two types of benchmarks, including both microbenchmarks and macrobenchmarks as shown in Figure~\ref{fig:benchmark}.

\begin{itemize}
\item \emph{Microbenchmarks} consist of a set of simple workloads targeting specific system resource usage, i.e., CPU-bound workloads, memory-bound workloads, diskIO-bound workloads, and network-bound workloads. \textbf{CPU-bound benchmarks} consist of two workloads. Specifically, \textit{sls-matrixMul} in Figure~\ref{fig:benchmark} calculates the result of the multiplication of two N-dimensional square matrices, and \textit{sls-linpack} solves the linear equation. These benchmarks will consume more computation power. \textbf{Memory-bound benchmarks} have a workload to calculate Fibonacci~\cite{fibonaccinew} values recursively, which will consume a lot of memory. \textbf{DiskIO-bound benchmarks} consist of three workloads, i.e., \textit{sls-dd}, \textit{sls-randomIO}, and \textit{sls-sequentialIO}. Specifically, \textit{sls-dd} is a workload that creates files in ``/tmp/'' directory of the local disk through using \textit{dd} command of the Linux system. \textit{sls-randomIO} and \textit{sls-sequentialIO} are workloads that measure the throughput and latency of random IO and sequential IO, respectively. \textbf{Network-bound benchmarks} also consist of three workloads. \textit{sls-http} is a network-bound test, which will be returned immediately after the invocations with a small JSON-format payload. \textit{sls-http} can be used to verify the round-trip time of different geographically distributed deployments. \textit{sls-iPerf} is designed to actively measure the maximum bandwidth achievable on the IP network. \textit{sls-cloudstorage} is a benchmark to measure the throughput and latency between function instances and cloud storage.

\item \emph{Macrobenchmarks} focus on real-world workloads that will consume multiple system resources (i.e., multimedia data process, MapReduce, machine-learning-based serving, etc.), and they go far beyond microbenchmarks. \textbf{Multimedia data process}: \textit{sls-image} and \textit{sls-video} workloads are designed in this kind of application. \textit{sls-image} is an image processing workload, which performs image transformation tasks using the Python Pillow library\cite{Pillow}. \textit{sls-image} fetches an input image from the cloud storage and applies ten different effects (e.g., copy, rotation, cropping, etc.) to this image. The corresponding output is uploaded back to the cloud storage. \textit{sls-video} applies the gray-scale effect from the OpenCV library\cite{OpenCV} to the video input and uploads the converted video to the cloud storage. \textbf{MapReduce}: MapReduce is a popular programming model that allows developers to process or generate large-scale data in parallel. We add a \textit{sls-mapreduce} workload, which consists of two types of functions. One is the \textit{Map} function to implement functionality filtering and sorting, and the other is the \textit{Reduce} function to merge and organize the outputs of the \textit{Map} function. \textbf{ML Training \& Serving}: Machine learning (ML) workloads mainly involve ML model training and ML model serving. Generally, the raw input data of a machine learning task needs pre-processing to prepare the input of training, e.g., \textit{sls-lr-training} workload in our benchmarks. We use the text dataset about \textit{Amazon Fine Food Review}\footnote{\url{https://snap.stanford.edu/data/web-FineFoods.html}} saved in the cloud storage, and input them into the regression model. Because the \textit{sls-lr-training} workload needs to access large-size datasets from the cloud storage, it often takes up a lot of CPU, memory, and network. After training a model, this model needs to be served for arbitrary inputs to make predictions, e.g., \textit{sls-lr-serving} workload that inputs users’ review texts into this model to predict the corresponding sentiment score. To further explore the inference related to deep learning models, we add the image classification workload named \textit{sls-cnn} and words generation workload named \textit{sls-rnn}. \textit{sls-cnn} uses a SqueezeNet model\cite{SqueezeNet} to achieve an impressive accuracy on an ImageNet\cite{ImageNet} with 50x fewer parameters than the state-of-the-art model. \textit{sls-cnn} is implemented with Python TensorFlow Keras\cite{TensorflowKeras}. Due to the limited memory size, attempts to import other convolutional neural network models (CNN\cite{CNN}) are proved to be a failure. \textit{sls-rnn} uses a recurrent neural network model (RNN\cite{RNN}) to implement words generation through PyTorch\cite{PyTorch}.
\end{itemize}

In addition, \testbedName also constructs a collection function to obtain the underlying runtime information for serverless functions to further analyze potential causes influencing concurrency performance. The runtime information mainly leverages the \emph{proc} filesystem on Linux (\emph{procfs}) that can expose global statistics of the underlying function instances and VMs, as well as other information like CPU, memory, and so on.

\section{Evaluation of the actual runtime performance}\label{sec:evaluation}


In our experiments, by default, we repeat each round test for 20 executions and use the median value to represent related evaluation results. The regions for functions were us-east-1, us-east1, eastus, us-east-1 in AWS Lambda, Google Cloud Functions, Azure Functions, and Alibaba Cloud Function Computing, respectively. Meanwhile, in our experiments, we also report the start time and end time of invocations, and function configurations for each round to facilitate further analysis. Most of our measurements have been done from June 2020 to March 2021. Note that we do not show all results in this paper due to the space limit. Some key results are shown as follows.

\subsection{Startup Latency}\label{sec:startuplatency}


The cold start of serverless applications may involve launching a new VM and function instance, downloading dependent packages, setting up the runtime environment, and initializing function, and these processes take more time to handle a request than reusing an existing function instance (i.e., \emph{warm} start). In the condition of cold start, it can significantly affect application responsiveness and in turn, affect the user experience. Thus, reducing cold start time is a key challenge in serverless computing~\cite{OakesATC18, JonasCoRR2019, DuASPLOS2020}. In our study, we focus on the overall cold start time instead of the cold start time of a certain sub-process. Following the previous work~\cite{WangATC2018}, we create functions with the same workload and configuration for each serverless computing platform and sequentially invoke them twice to derive the overall cold start time. Specifically, the difference of overall response time, i.e., end-to-end duration from the client perspective, is considered as an estimation of its cold start latency. We will investigate how programming languages, memory size, and package size affect the cold start time of applications.


\subsubsection{Programming languages}

\begin{figure}[htbp]
\centering
	\subfigure[Python]{
		\begin{minipage}[b]{0.32\textwidth}
			\includegraphics[width=\textwidth]{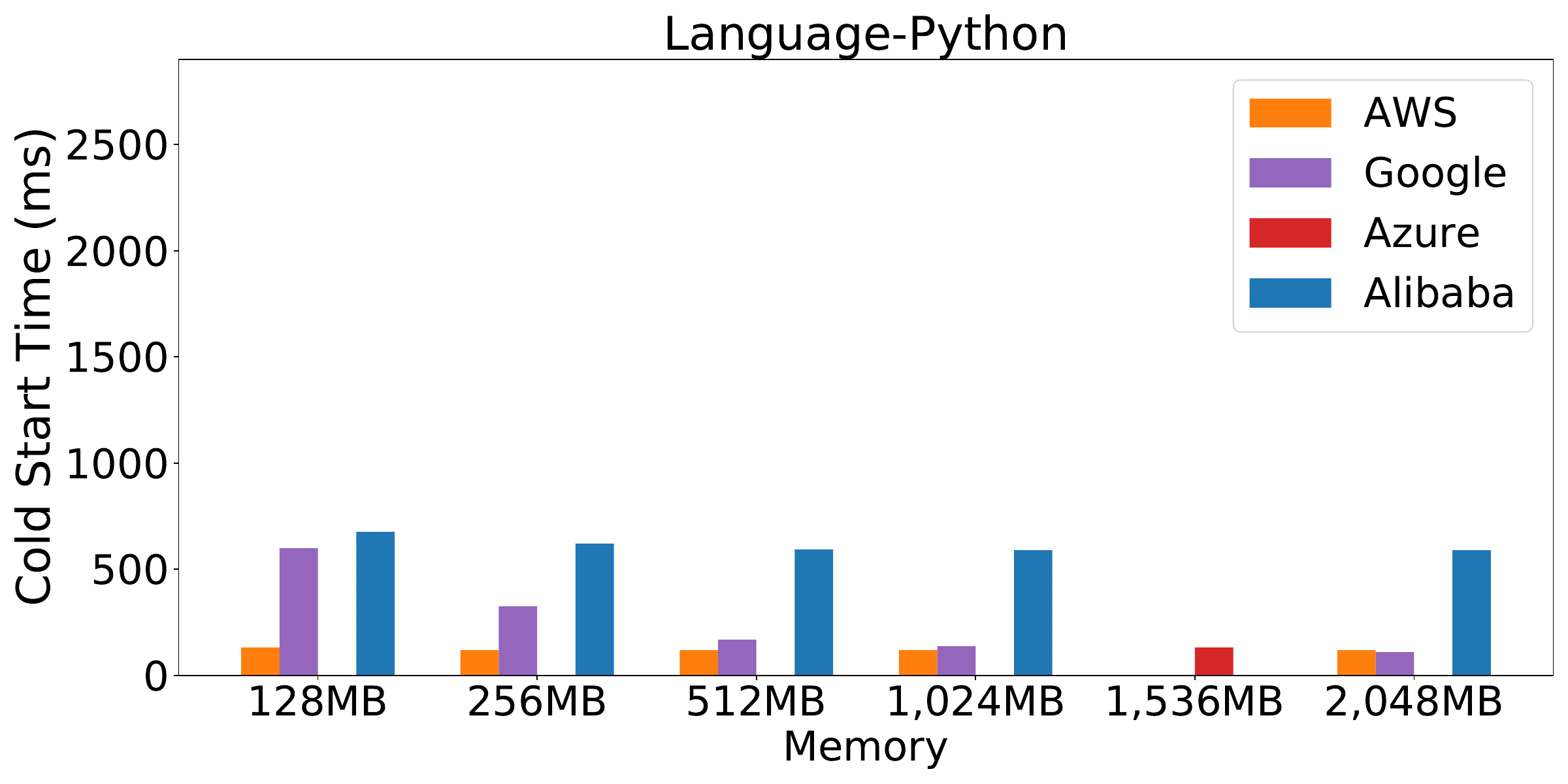}
		\end{minipage}
		\label{fig:lan-python}
	}
    	\subfigure[Node.js]{
		\begin{minipage}[b]{0.32\textwidth}
			\includegraphics[width=\textwidth]{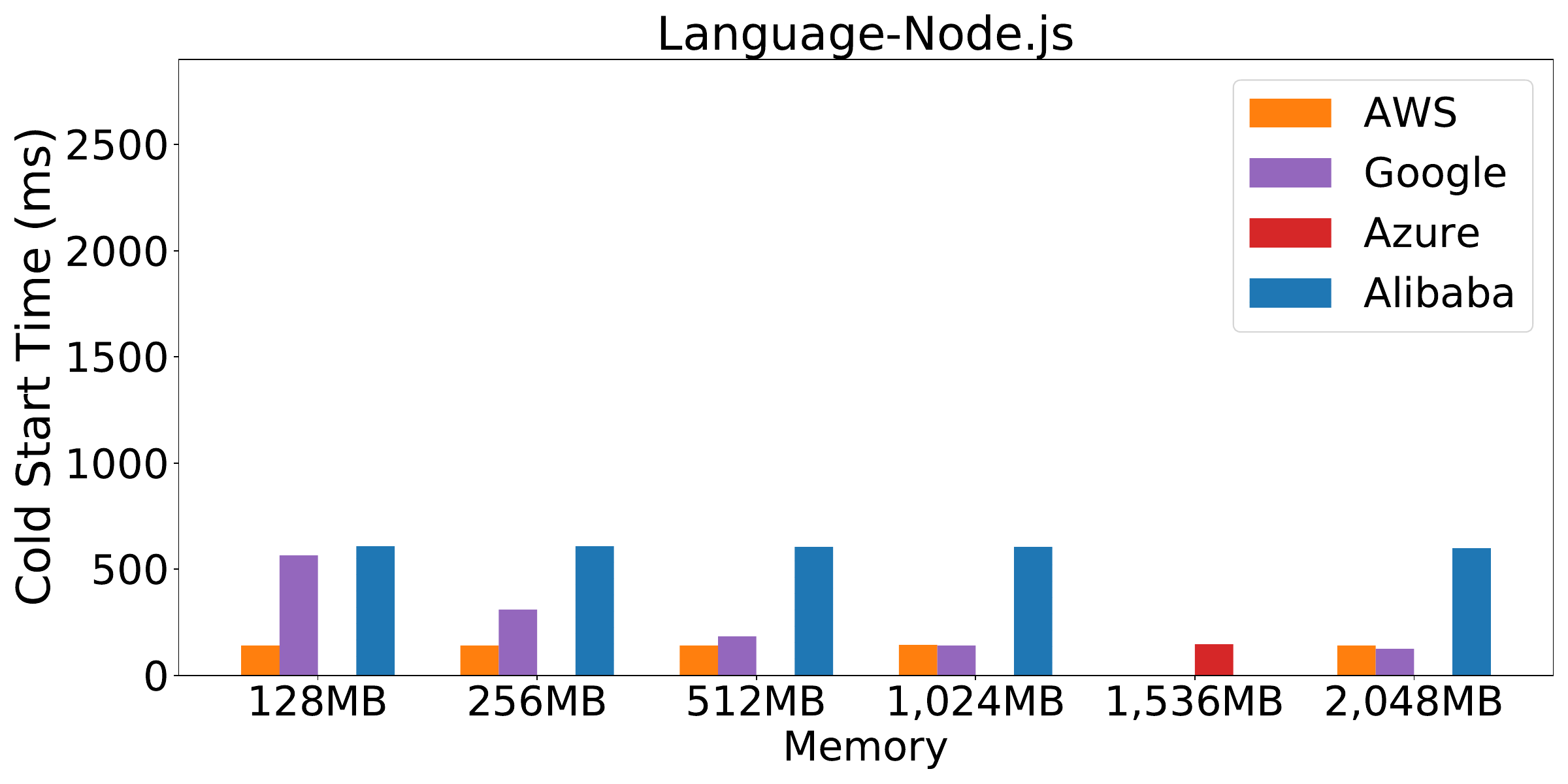}
		\end{minipage}
		\label{fig:lan-nodejs}
	}
    	\subfigure[Java]{
		\begin{minipage}[b]{0.32\textwidth}
			\includegraphics[width=\textwidth]{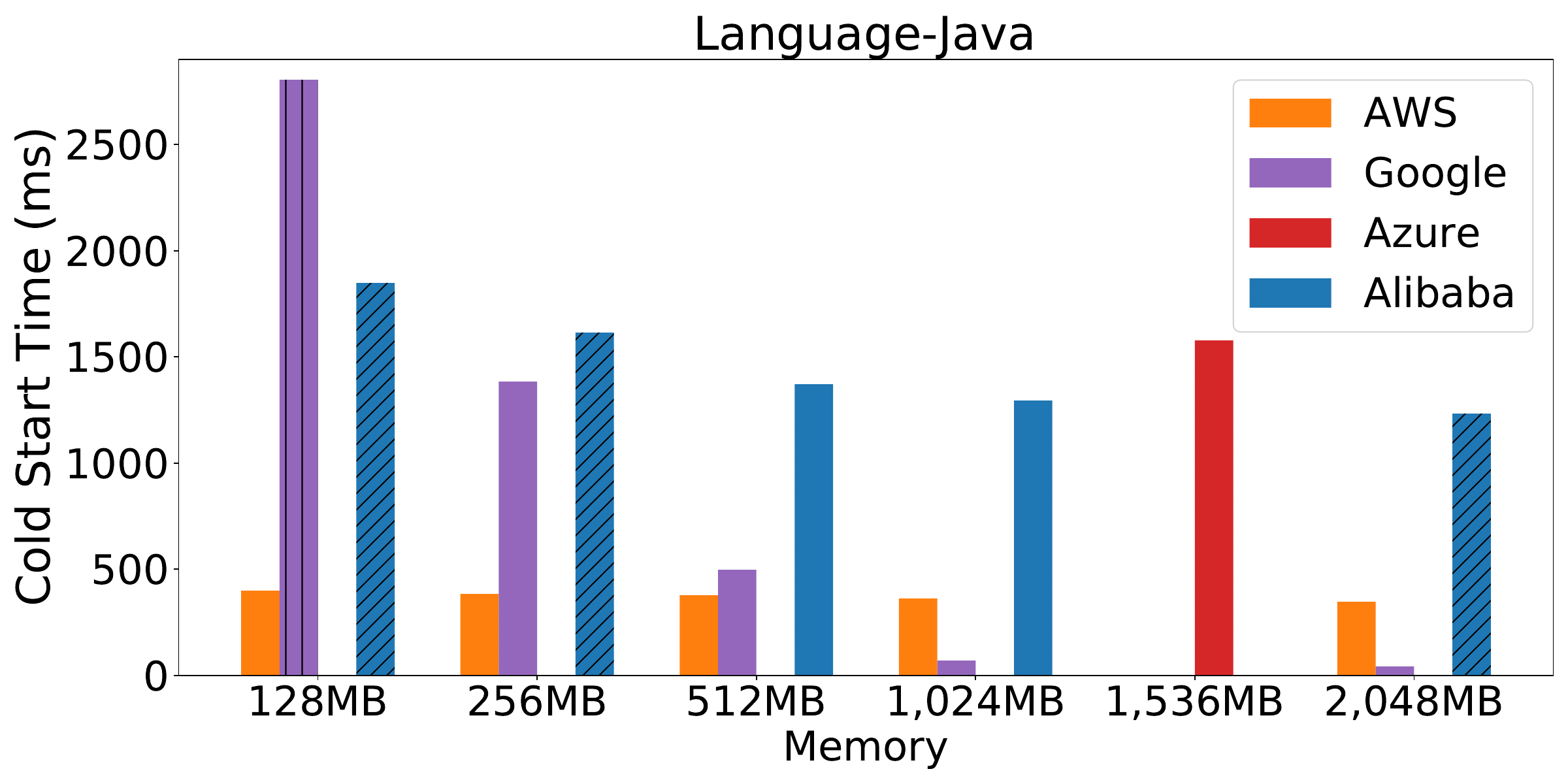}
		\end{minipage}
		\label{fig:lan-java}
	}
    \caption{The distribution of the cold start time with different programming languages.}
    \label{fig:lan-python_nodejs_java}
\end{figure}




As shown in Figure~\ref{fig:language}, we find that programming languages are not well supported across all platforms. Thus, we mainly compare the cold start time of the three most well-supported languages\cite{toplanguagenew}, i.e., Python, Node.js, and Java. Figure~\ref{fig:lan-python}, Figure~\ref{fig:lan-nodejs}, and Figure~\ref{fig:lan-java} show the distribution of the cold start time for three different languages, respectively. The most obvious trend is that statically typed languages (e.g., Java) have over 3 times higher cold start latency than dynamically typed language (e.g., Python and Node.js), especially for the cases with small memory. AWS Lambda is overall the fastest with allocated memory less than 1,024 MB. In 512 MB memory, AWS Lambda has a median cold start latency of only 228.37 ms for Python, 139.96 ms for Node.js, and 374.79 ms for Java. Executing a Java function needs to launch and initiate a ``cumbersome'' JVM, resulting in much more overhead and cold start time. By contrast, launching a Java function on AWS Lambda has the least cold start time than other platforms. When allocating 128 MB memory, AWS Lambda even speeds up by a factor of about 7 compared to Google Cloud Functions. Interestingly, we also find that the Google Cloud Functions can perform better than other platforms to run Java functions with allocated memory of more than 1,024 MB, and its cold start time can even be less than functions developed by other languages.

\noindent\textbf{Discussion and implications.} Regardless of preference, developers can choose ``light-weight'' languages (e.g., Python, Node.js) to develop their functions with lower cold start time, especially for time-sensitive tasks.  When developing functions with ``big'' memory (e.g., more than 1,024 MB), developers are free to choose any language with low cold start latency. For cloud vendors, they can further optimize the initialization of the cumbersome runtime like Java. For example, cloud vendors can apply new technologies like Unikernel to reduce the overhead of initialization\cite{JonasCoRR2019}.

\subsubsection{Memory sizes}

As shown in Figure~\ref{fig:info}, developers specify the function memory when deploying their functions, and serverless computing platforms will allocate CPU capability proportional to the allocated memory. Therefore, allocating more memory to functions will also increase the computing power of function instances. Since we cannot specify the memory of Azure functions (1,536 MB at most), we can assume that its performance is between 1,024 MB and 2,048 MB. In addition, according to Figure~\ref{fig:info}, we find that Google Cloud Functions supports fixed memory sizes, i.e., 128, 512, 1,024, 2,048, and 4,096 MB. Thus, we design the comparison of the other three platforms under these memory sizes. As shown in Figure~\ref{fig:lan-python}, Figure~\ref{fig:lan-nodejs}, and Figure~\ref{fig:lan-java}, we find that more memory allocated to function results in the lower cold start time for most cases for Google Cloud Functions. Meanwhile, \textbf{its memory size reduces the cold start time in a roughly linear fashion}. 98.5\% of the cold start time of Java is reduced on Google Cloud Functions when increasing the memory from 128 MB to 2,048 MB. However, for AWS Lambda and Alibaba Cloud Function Compute, they do not seem to have significant differences with different allocated memories for Python and Node.js applications. It probably means that not so much memory and CPU are required to launch these applications.

\noindent\textbf{Discussion and implications.} For functions written with Python or Node.js on AWS Lambda and Alibaba Cloud Function Compute, allocating more memory will not speed up the cold start time and only lead to more cost. Developers do not need to increase memory allocation if the task is not memory-intensive. In other cases, developers can properly increase the memory size to reduce the cold start time. For cloud vendors, they can give some practical guides of memory configuration to make a balance between the cost and performance under different program languages.

\subsubsection{Package sizes}

\begin{figure}[htbp]
\centering
\begin{minipage}[t]{0.48\textwidth}
\centering
\includegraphics[width=\textwidth]{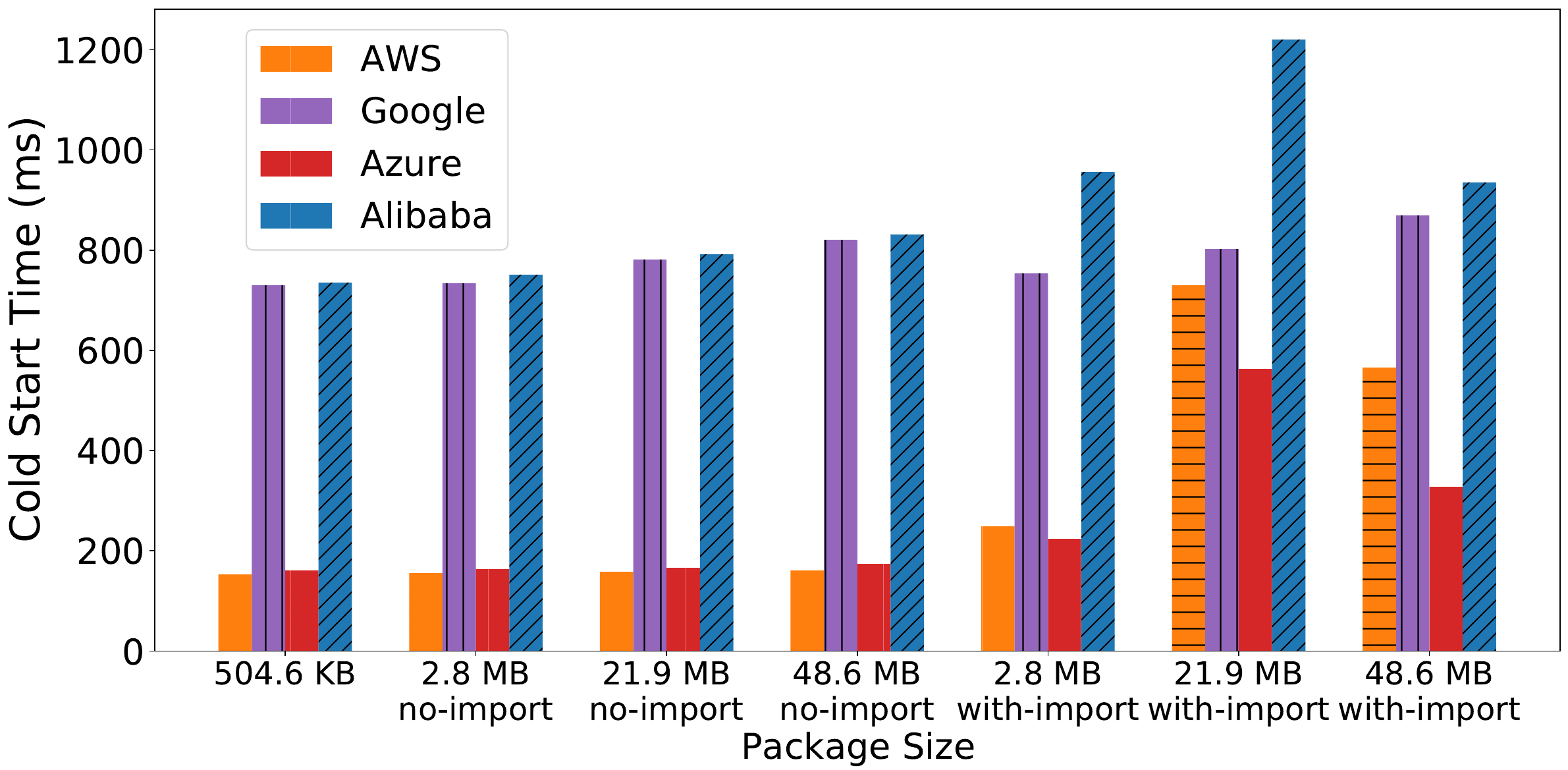}
\caption{The distribution of cold start time of different package sizes.}
\label{fig:cold_start_package}
\end{minipage}

\end{figure}

Figure~\ref{fig:cold_start_package} compares four Python functions with 128 MB of memory under the various number of third-party packages. The first group bar presents the cold start time of a basal function (with 504.6 KB). The following three group bars show the cold start time of functions, adding a third-parity library of different sizes into the basal function without loading (2.8 MB with a library named \textit{Pillow}, 21.9 MB with a library named \textit{Numpy}, and 48.6 MB with a library named \textit{OpenCV-python}, respectively). The last three group bars show the distribution of cold start time of functions with loading the useless library, i.e., the dependency libraries are redundantly imported in the deployment package. \textbf{We find that only increasing the size of the deployment package without loading does not obviously affect the cold start time on AWS Lambda and Azure Functions. Instead, loading more libraries in the function will dramatically increase the cold start time except on the Google Cloud Functions}, since it will increase the time of initializing functions (e.g., loading code into memory).  In particular, the cold start time can even increase 4.6x if the function loads some unused libraries (21.9 MB-with-import VS 21.9 MB-no-import). In the package size experiment, we select Python functions as an example to explore the impact of package size on the cold start time and provide a relevant snapshot about these serverless computing platforms. Certainly, leveraging our evaluation tool \testbedName, researchers can also reproduce experiments with other languages (e.g., JavaScript, Java).

\noindent\textbf{Discussion and implications.} Loading redundant libraries may introduce non-negligible cold start time, thus developers need to trim their code (e.g., removing the useless code and import) and apply lazy loading of libraries to alleviate the overhead of initializing functions. It is better for cloud vendors to provide related tools to help developers optimize their function packages before development instead of directly deploying the zipped function packages submitted by developers.

\subsection{Resource Efficiency}\label{sec:resourceefficiency}


We leverage our benchmarks to explore the resource efficiency of serverless computing platforms. First, we compare the performance of microbenchmarks (i.e., CPU-bound, memory-bound, diskIO-bound, and network-bound workloads) on different serverless computing platforms. Second, we use macrobenchmarks to explore complex resource utilization. In our study, we focus on metrics like the execution time of function, and the throughput of storage and CPU. 


\subsubsection{Microbenchmarks performance}

\begin{figure}[htbp]
\centering
\begin{minipage}[t]{0.48\textwidth}
\centering
\includegraphics[width=\textwidth]{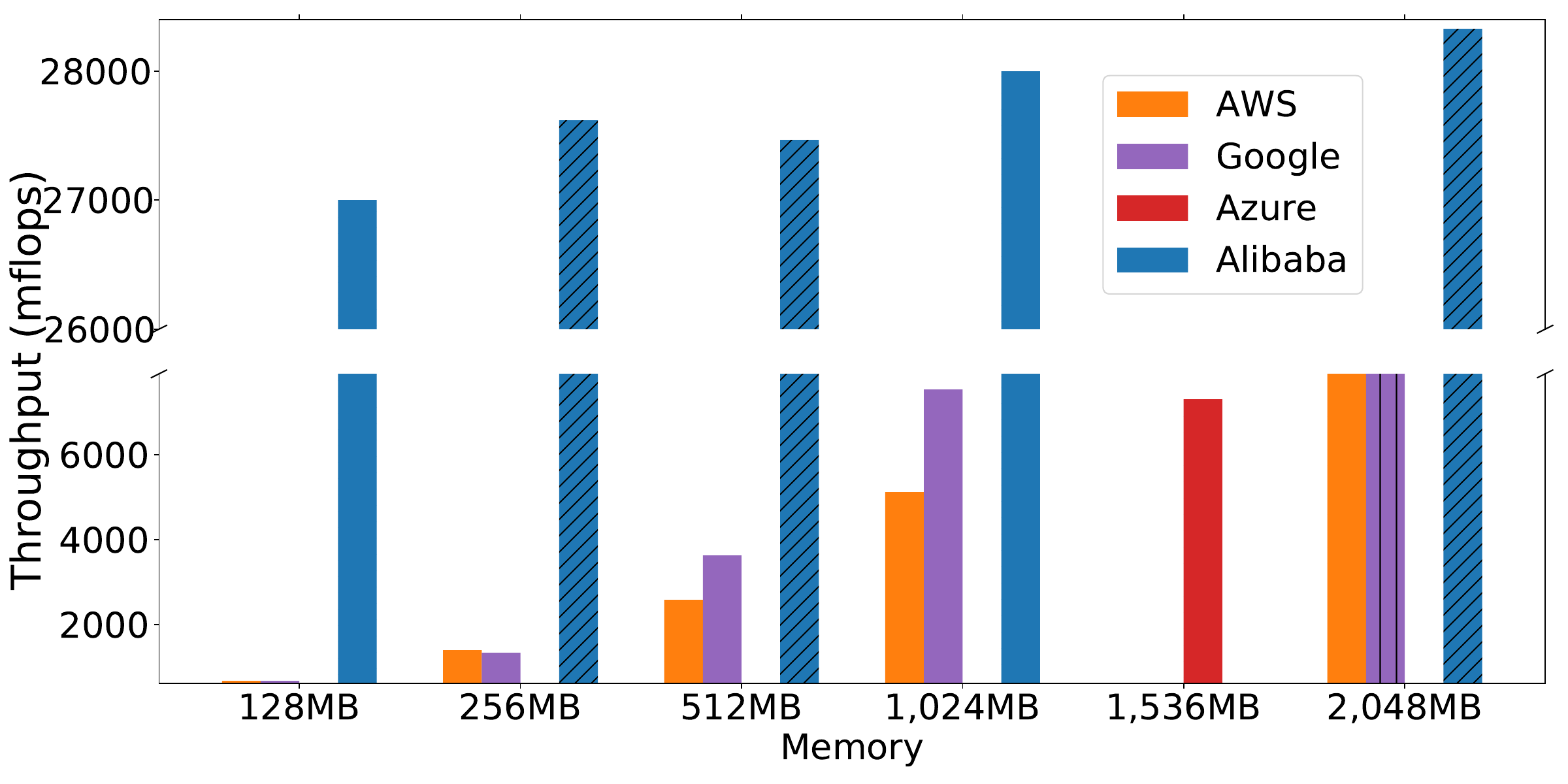}
\caption{The CPU throughput for the CPU-bound workload.}
\label{fig:linpack}
\end{minipage}
\begin{minipage}[t]{0.48\textwidth}
\centering
\includegraphics[width=\textwidth]{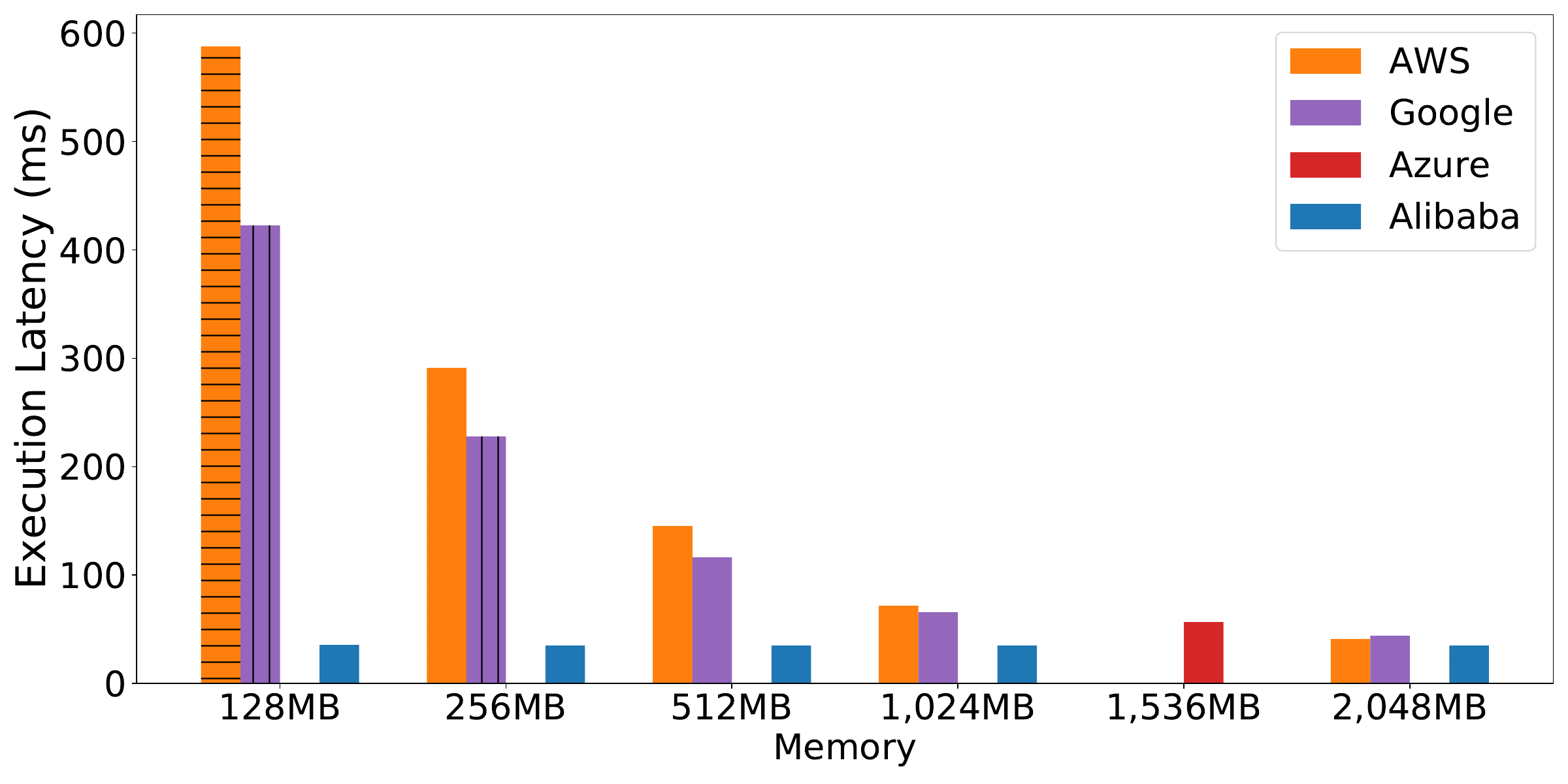}
\caption{The execution latency for the memory-bound workload.}
\label{fig:memory_bench}
\end{minipage}
\end{figure}


We use the \textit{sls-linpack} benchmark to evaluate the CPU capacity of different serverless computing platforms, and the results are shown in Figure~\ref{fig:linpack}. \textit{sls-linpack} operates on 1,000$\times$1,000 matrix and outputs a performance rating metric in terms of millions of floating-point operations per second (mflops). We find that Alibaba Cloud Function Compute performs much better than other platforms while AWS Lambda performs worst. In addition, for AWS Lambda and Google Cloud Functions, the \textit{mflops} can increase linearly by allocating more memory, conforming to the finding as we summarized in Figure~\ref{fig:info} that serverless computing platform will allocate CPU capability proportional with allocated memory.

To measure the performance of memory-bound workloads, we use the \textit{sls-fib} to calculate the 25th Fibonacci value recursively. This process will result in high memory consumption, and the results are shown in Figure~\ref{fig:memory_bench}. Alibaba Cloud Function Computing has the shortest execution time among all cases, and it is insensitive to more memory. For other platforms, allocating more memory can linearly reduce the execution time. However, the rate of reduction will decrease when allocated memory increases from 1,024 MB to 2,048 MB.






\begin{figure}[htbp]
\centering
\begin{minipage}[t]{0.48\textwidth}
\centering
\includegraphics[width=\textwidth]{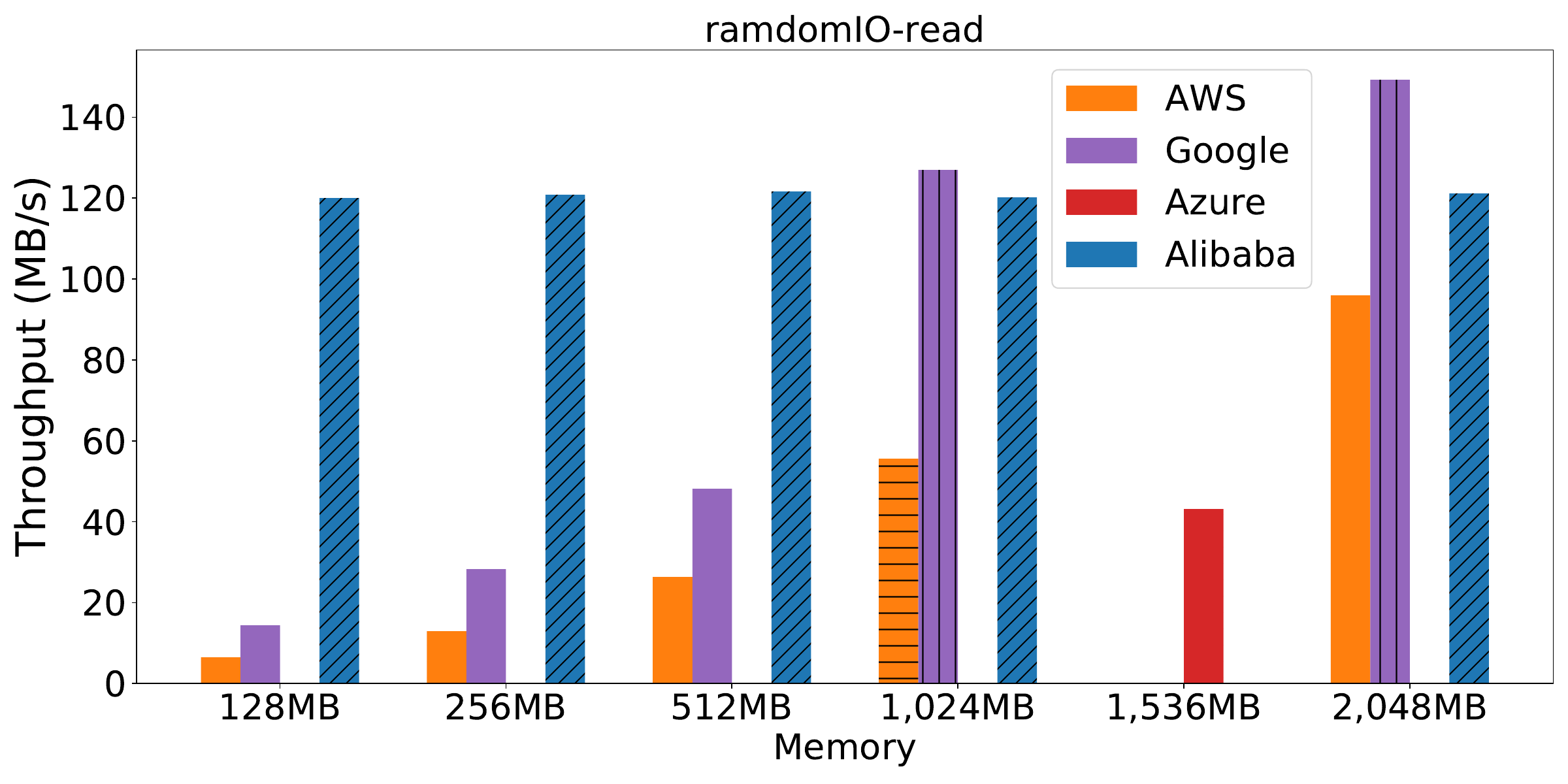}
\caption{The random read throughput.}
\label{fig:ramdomIO-read}
\end{minipage}
\begin{minipage}[t]{0.48\textwidth}
\centering
\includegraphics[width=\textwidth]{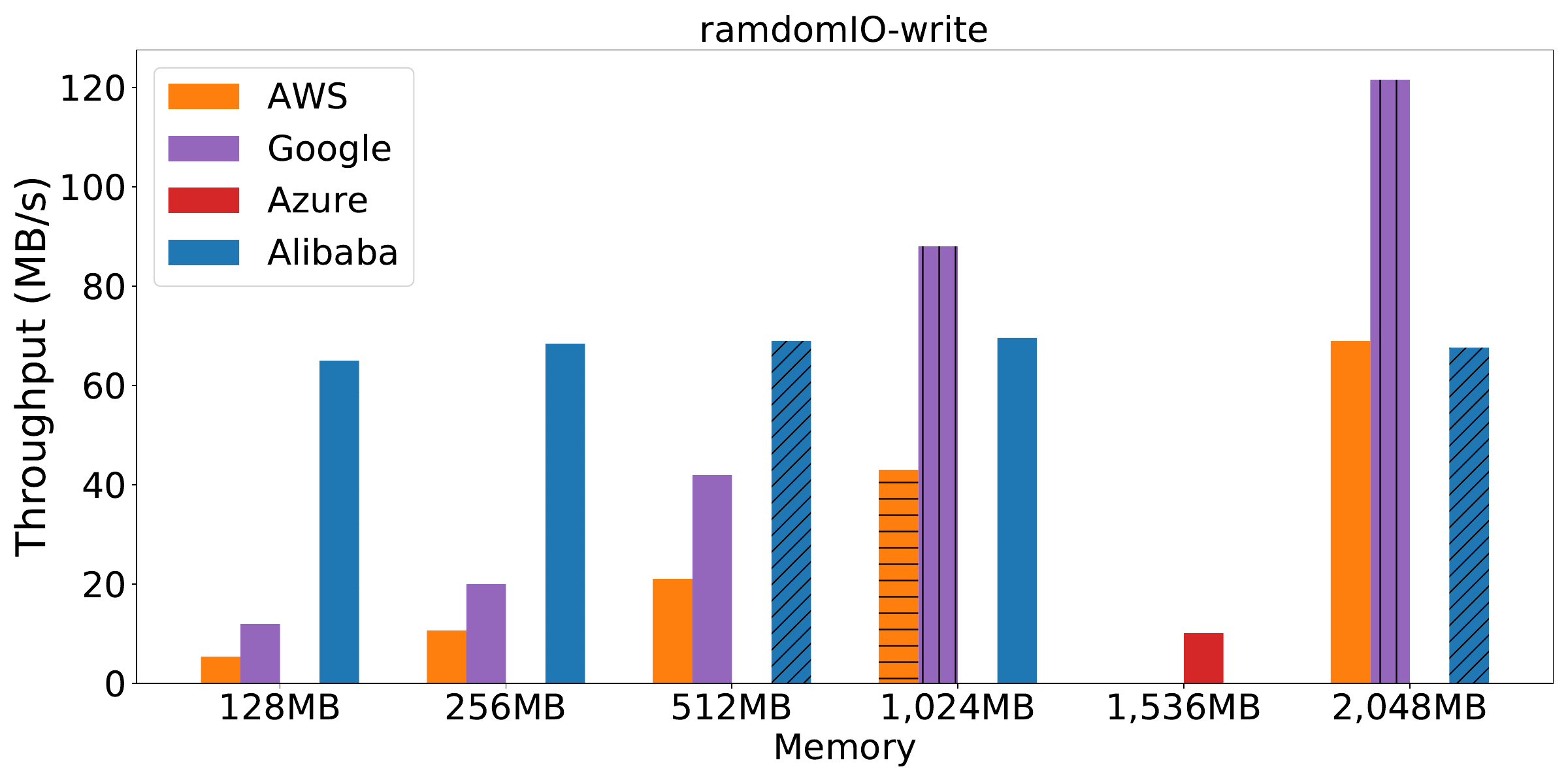}
\caption{The random write throughput.}
\label{fig:ramdomIO-write}
\end{minipage}
\end{figure}



\begin{figure}[htbp]
\centering
\begin{minipage}[t]{0.48\textwidth}
\centering
\includegraphics[width=\textwidth]{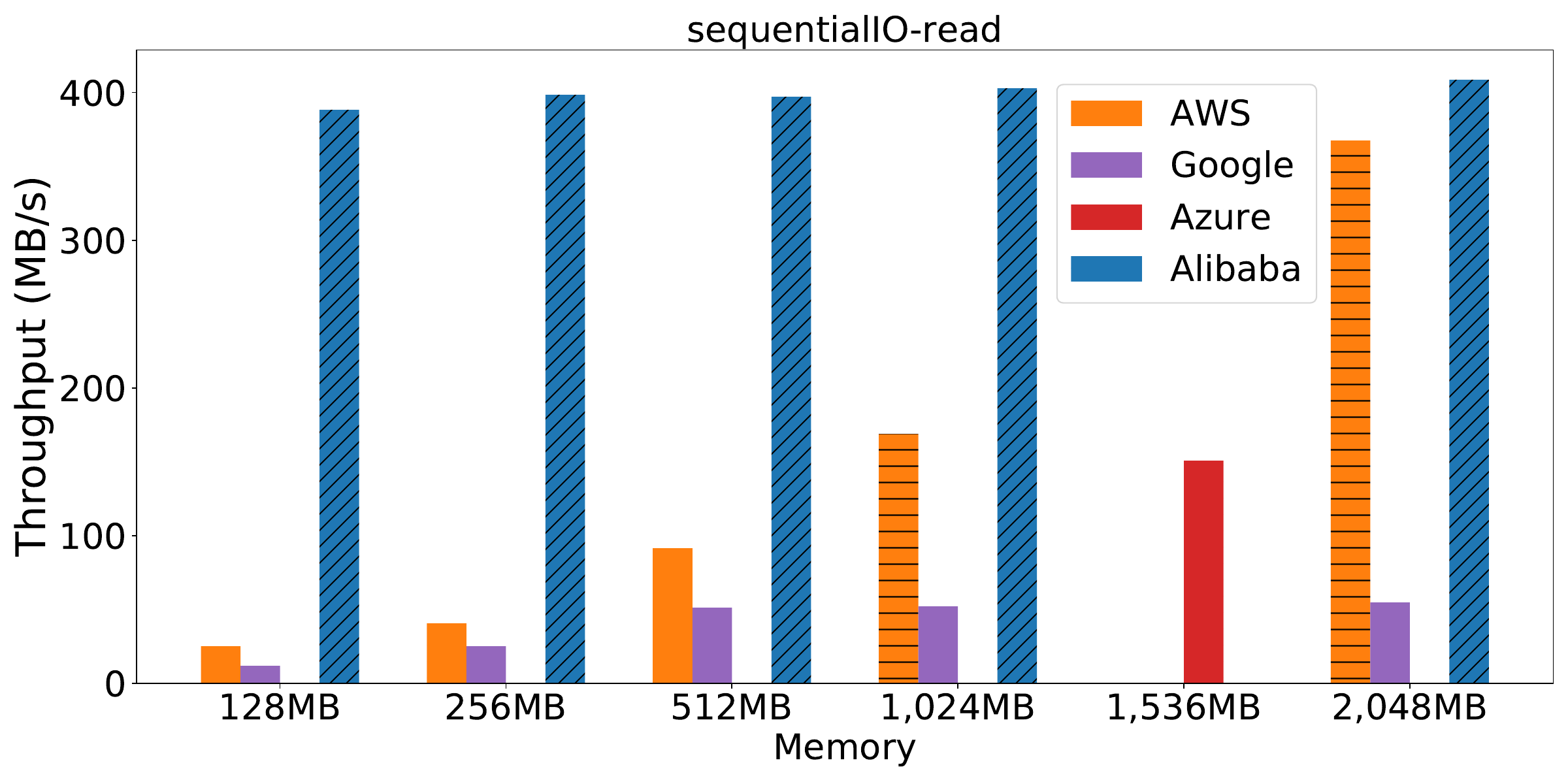}
\caption{The sequential read throughput.}
\label{fig:sequentialIO-read}
\end{minipage}
\begin{minipage}[t]{0.48\textwidth}
\centering
\includegraphics[width=\textwidth]{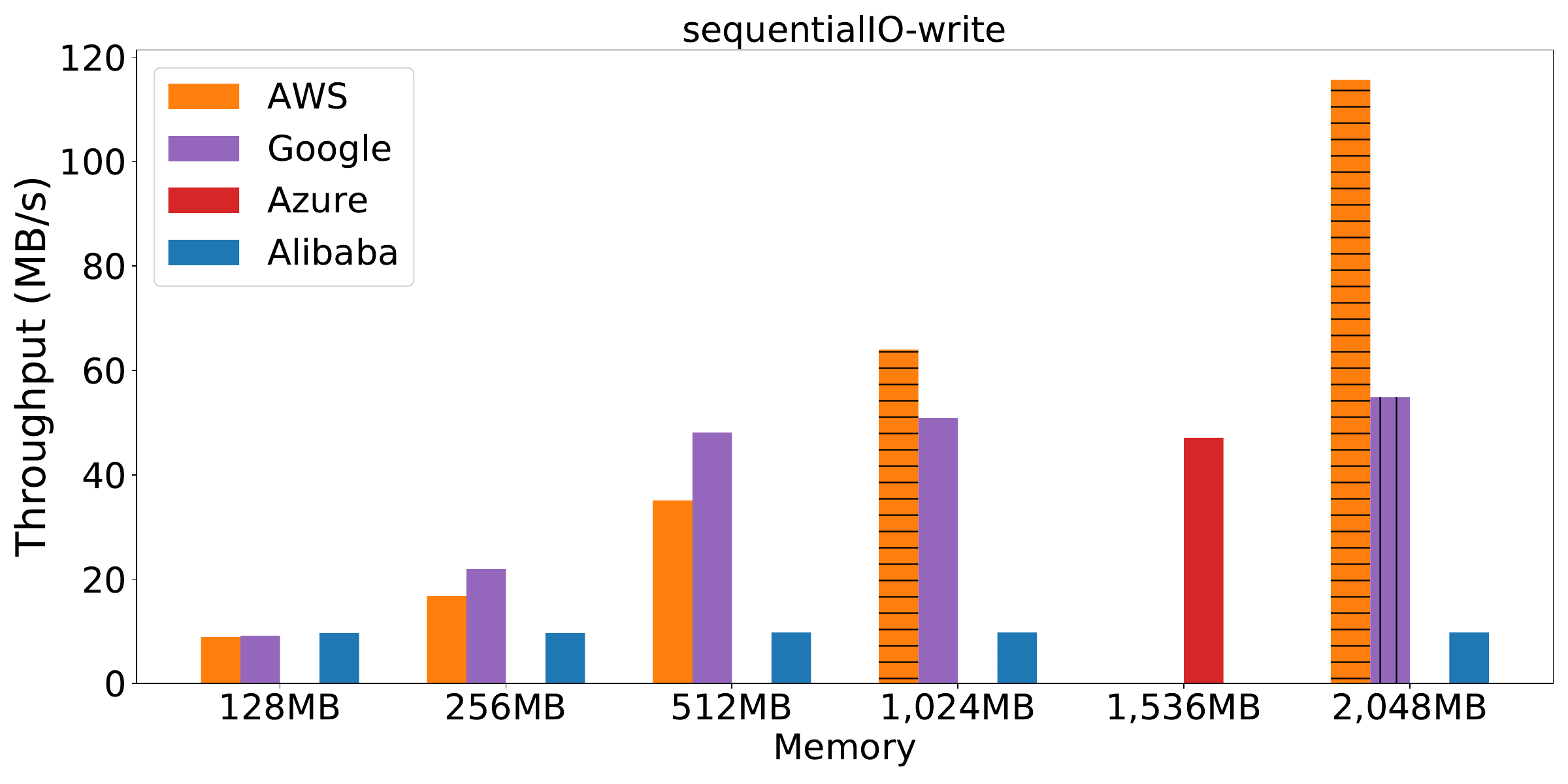}
\caption{The sequential write throughput.}
\label{fig:sequentialIO-write}
\end{minipage}
\end{figure}



We leverage \textit{sls-randomIO} and \textit{sls-sequentialIO} to measure the throughput of local storage in terms of different IO patterns, i.e., read and write. From the results of Figures~\ref{fig:ramdomIO-read},~\ref{fig:ramdomIO-write},~\ref{fig:sequentialIO-read} and~\ref{fig:sequentialIO-write}, we observe that the IO throughput of AWS Lambda and Google Cloud Functions increases as larger memory is allocated. In random IO operations, Google Cloud Functions is better than AWS Lambda, while they are opposite in sequential IO operations. In addition, Alibaba Cloud Function Compute has not only a stable and relatively efficient performance under different memories but also an outstanding sequential IO read throughput bandwidth than other platforms shown in Figure~\ref{fig:sequentialIO-read}. Meanwhile, in 128 MB of memory, Alibaba Cloud Function Compute has 32.96 times the read throughput of Google Cloud Functions. Next, we analyze some specific cases. The IO throughput of Alibaba Cloud Function Compute performs poorly in the sequential write tasks. The reason is that sequential writing operators cause more IO competition. The sequential IO throughput of Google Cloud Functions increases by allocating memory, but the rate of increase will drop when exceeding a threshold. For example, sequential IO throughput of Google Cloud Functions trends to be stable when allocating memory more than 512 MB shown in Figures~\ref{fig:sequentialIO-read} and~\ref{fig:sequentialIO-write}. In addition, Google Cloud Functions performs better than Alibaba Cloud Function Compute in random read and write when allocated memory is more than 1024 MB shown in Figures~\ref{fig:ramdomIO-read} and \ref{fig:ramdomIO-write}. Surprisingly, we find that the random IO throughput of Azure Functions is lower than 1,024 MB instances of other platforms.



Since the network latency heavily depends on network condition and geographical location, we mainly focus on the throughput measurement with \textit{sls-iPerf} in this paper. \textit{sls-iPerf} uses \textit{iPerf3}\cite{iPerf3} with default configurations to run the throughput test for 30 seconds with the same-region iPerf servers, so that iPerf server-side bandwidth is not a bottleneck\cite{WangATC2018}. Due to the changeable network conditions, we cannot get stable results and just show some key findings in this part. Network throughput has a trend to increase as function memory increases. Actually, we have also measured the latency with the \textit{sls-http} benchmark deployed in different regions, and the latency indeed varies across different regions.

\noindent\textbf{Discussion and implications.}  Developers should choose the appropriate serverless computing platform to obtain better performance based on their resource needs of applications. For computation-intensive and memory-intensive workloads, the processing ability of each serverless computing is different. Developers can avoid the effect of memory on performance, and Alibaba Cloud Function Compute is recommended. However, for other platforms, developers can obtain more computing power to execute their computation-intensive and memory-intensive workloads by allocating more memory, but they need to make a trade-off between performance and cost.  If developers are deploying applications with frequent random IO operations, it is better to deploy them on Alibaba Cloud Function Compute, especially for cases with small allocated memory. When developers have decided to adopt AWS Lambda or Google Cloud Functions, they are advised to use AWS Lambda in sequential IO tasks, and Google Cloud Functions in random IO tasks.

\subsubsection{Macrobenchmarks performance}

\begin{figure}[htbp]
\centering
\begin{minipage}[t]{0.48\textwidth}
\centering
\includegraphics[width=\textwidth]{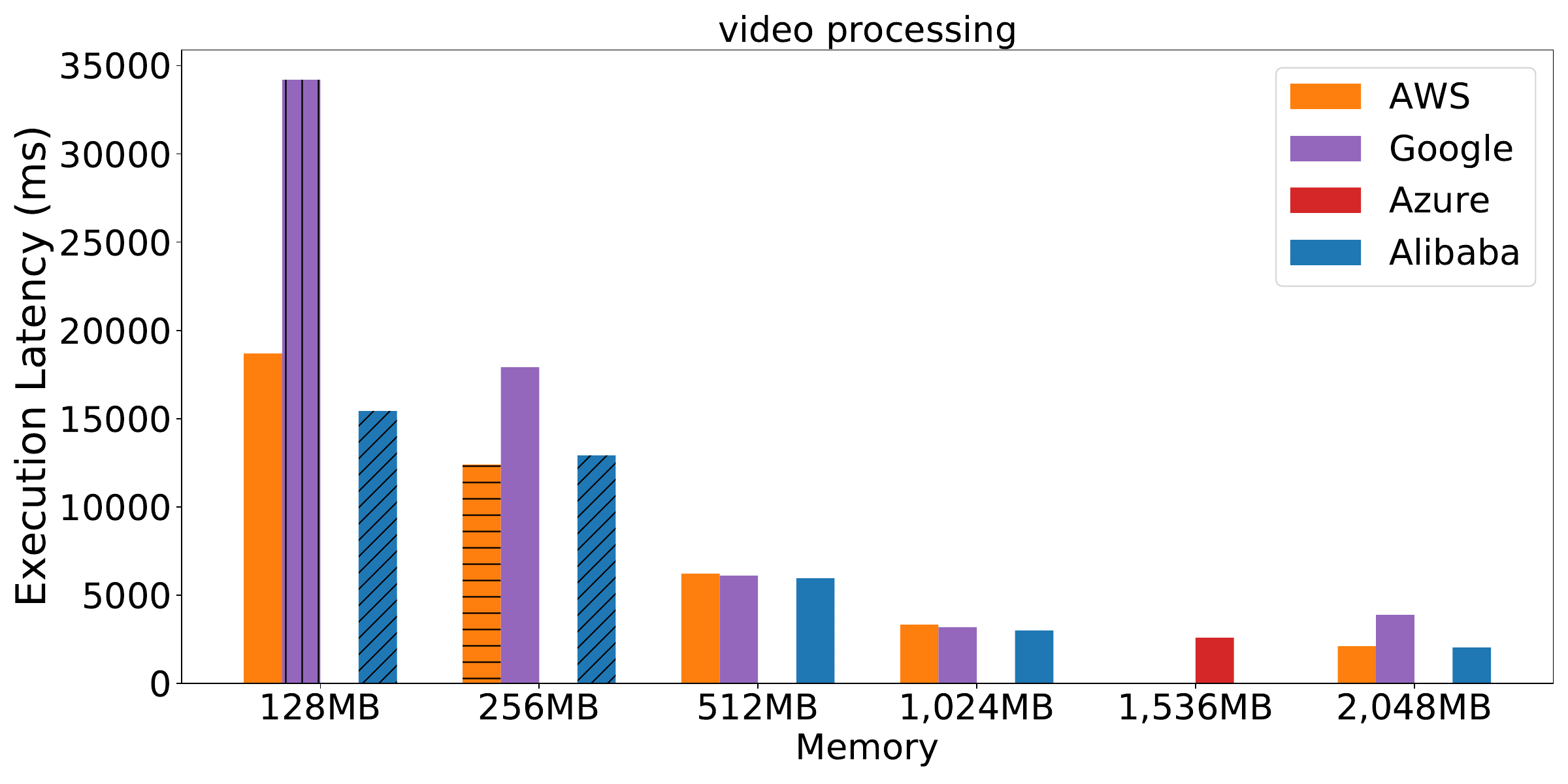}
\caption{The execution latency of \textit{sls-video}.}
\label{fig:video}
\end{minipage}
\begin{minipage}[t]{0.48\textwidth}
\centering
\includegraphics[width=\textwidth]{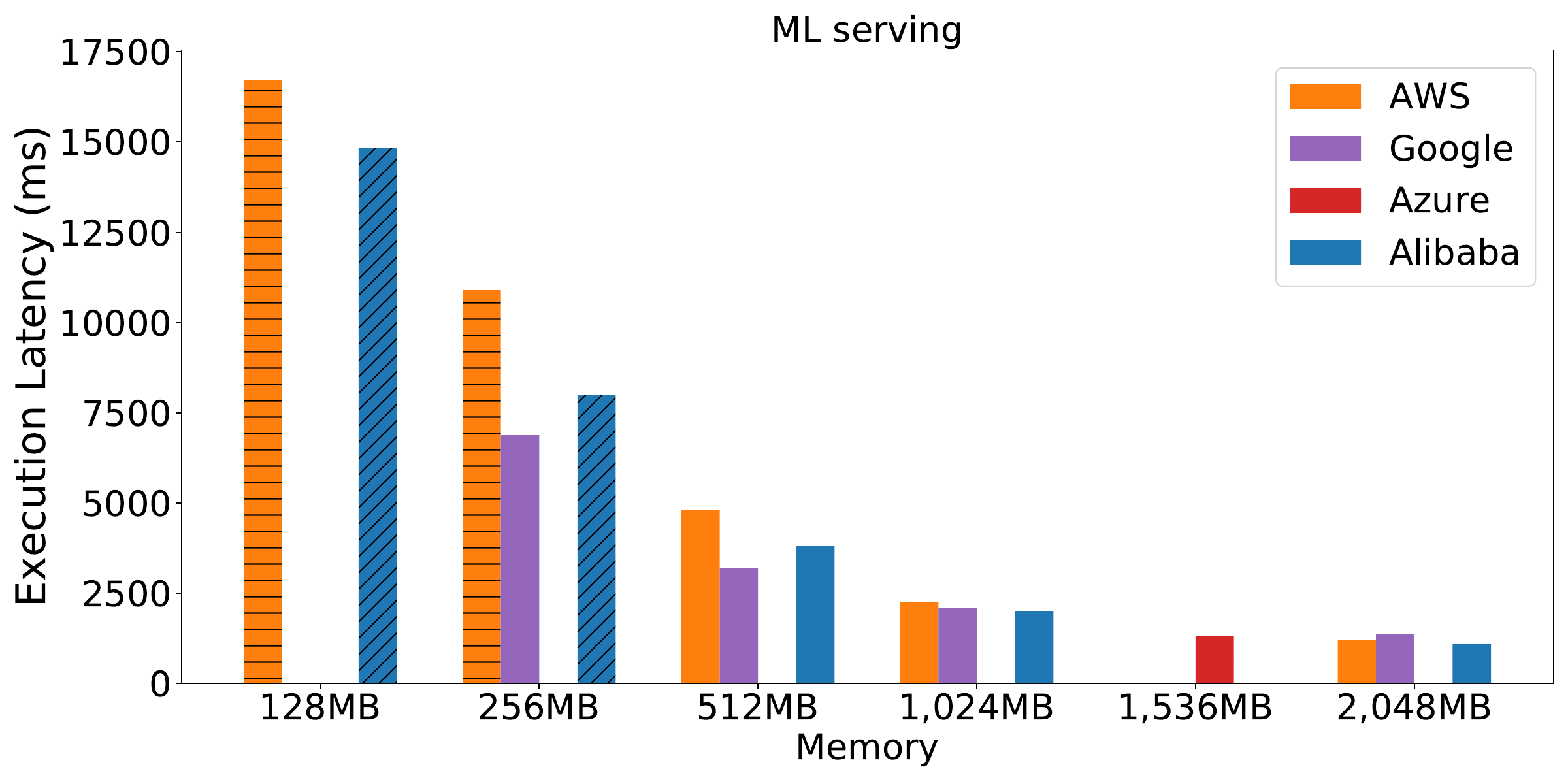}
\caption{The execution latency of \textit{sls-lr-serving}.}
\label{fig:serving}
\end{minipage}
\end{figure}



Different from microbenchmarks that exclusively evaluate different resources, macrobenchmarks utilize CPU, memory, disk IO, network resources, and so on together at different degrees. Figures~\ref{fig:video} and~\ref{fig:serving} show the execution latency to complete the video processing task (\textit{sls-video}) and model serving task (\textit{sls-lr-serving}) on each serverless computing platform, respectively. In this experiment, we also vary the allocated memory size of the function to explore the impacts on the execution latency of such macrobenchmarks. Missing bars in the figures indicate that the corresponding platform could not complete the given workload with the allocated memory size. From Figures~\ref{fig:video} and~\ref{fig:serving}, we find that allocating more memory can linearly reduce the execution time of functions, and narrow the performance gap among different serverless computing platforms. In addition, Alibaba Cloud Function Compute shows better performance than other platforms overall with the same allocated memory, confirming our previous findings of evaluations with microbenchmarks. In other words, Alibaba Cloud Function Compute performs well in most cases with microbenchmarks, thereby also has better performance for complicated workloads (i.e., macrobenchmarks). Although Azure Functions allocates 1,536 MB of memory, we find that the execution latency on Azure Functions is close to 1,024 MB instances of other platforms.

\noindent\textbf{Discussion and implications.} For those complicated workloads like macrobenchmarks, developers can allocate more memory for them to improve their performance. When memory is set as more than 1,024 MB, these serverless computing platforms can arrive at relatively stable performance.  Based on our results, Alibaba Cloud Function Compute has a better ability to handle such macrobenchmarks. 


 



\subsection{Concurrency Performance}\label{sec:scalability}

\begin{figure}[htbp]
\centering
\subfigure[AWS Lambda]{
		\begin{minipage}[b]{0.48\textwidth}
			\includegraphics[width=\textwidth]{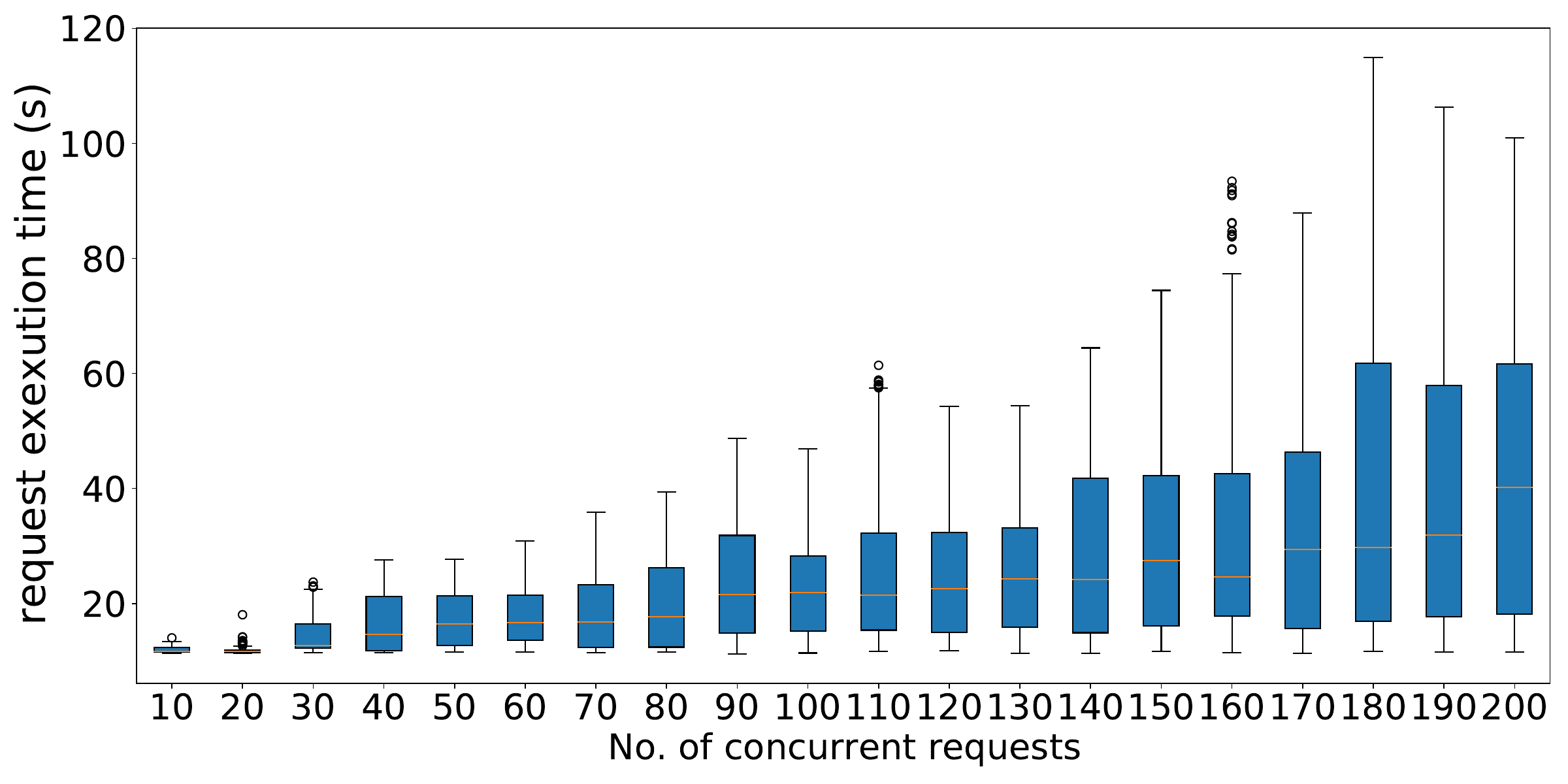}
		\end{minipage}
		\label{fig:aws-1024-performace}
}
\subfigure[Google Cloud Functions]{
		\begin{minipage}[b]{0.48\textwidth}
			\includegraphics[width=\textwidth]{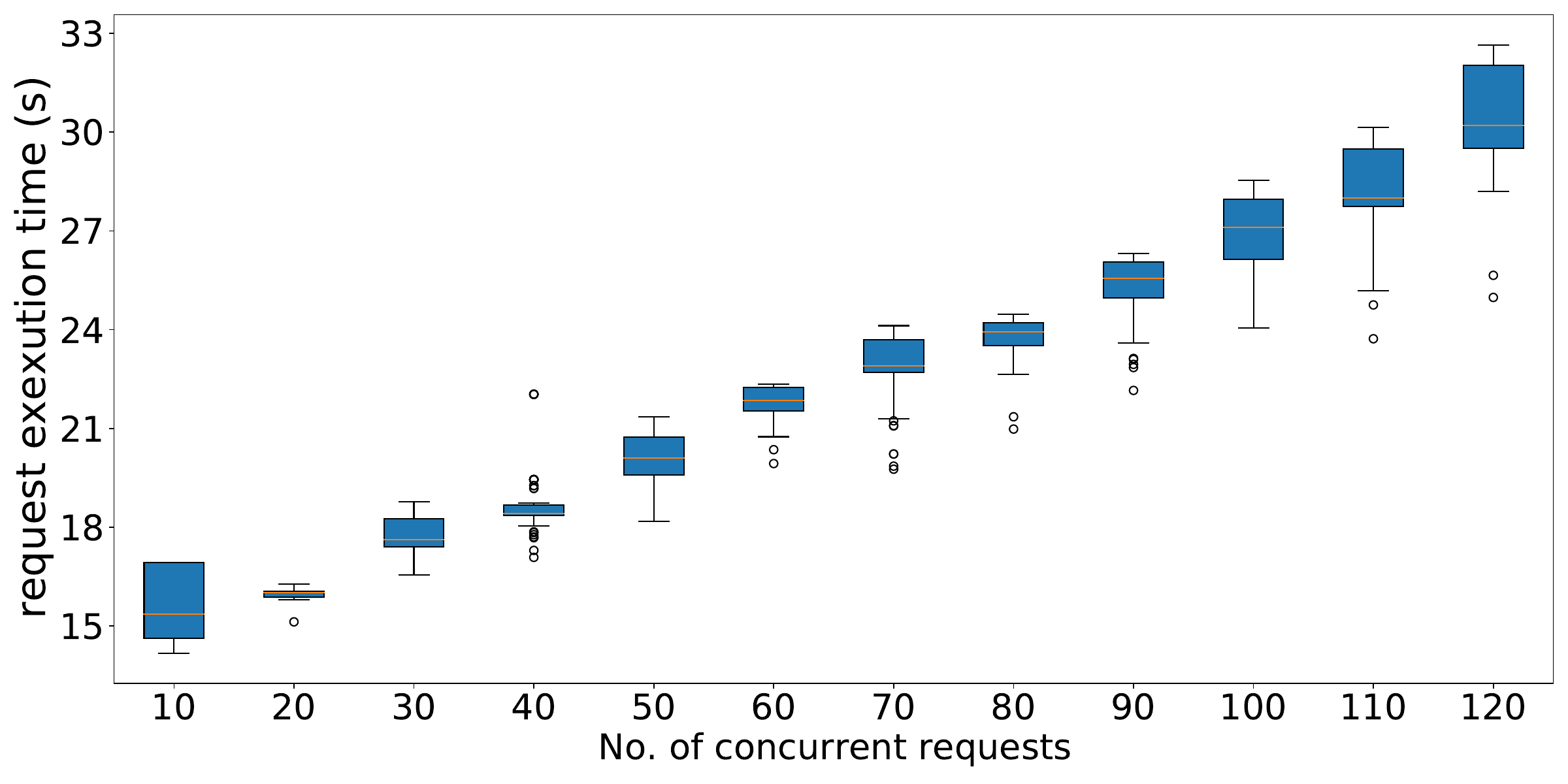}
		\end{minipage}
		\label{fig:google-1024-performace}
}
\subfigure[Alibaba Cloud Function Compute]{
		\begin{minipage}[b]{0.48\textwidth}
			\includegraphics[width=\textwidth]{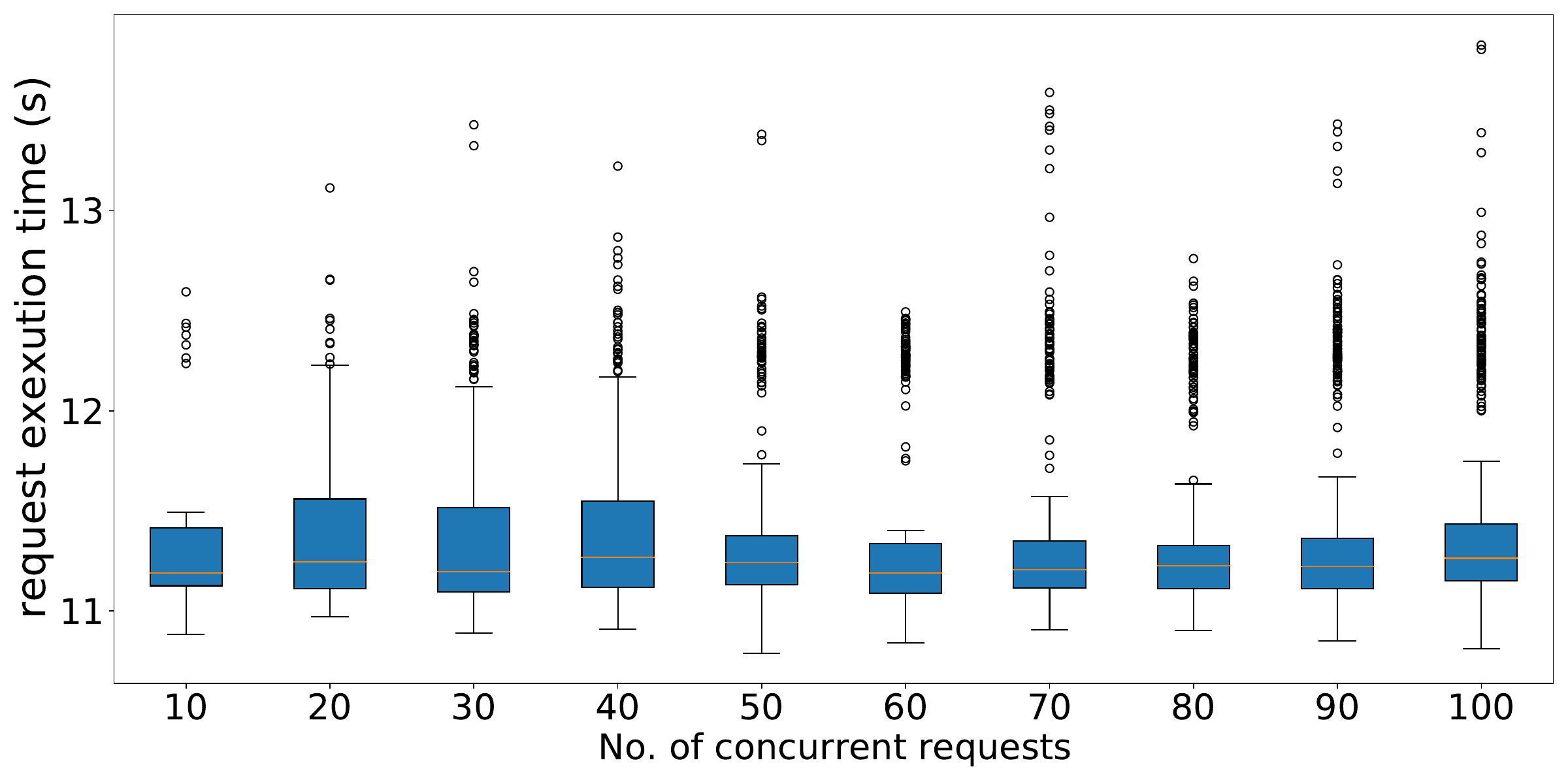}
		\end{minipage}
		\label{fig:ali-1024-performace}
}
\subfigure[Azure Functions]{
		\begin{minipage}[b]{0.48\textwidth}
			\includegraphics[width=\textwidth]{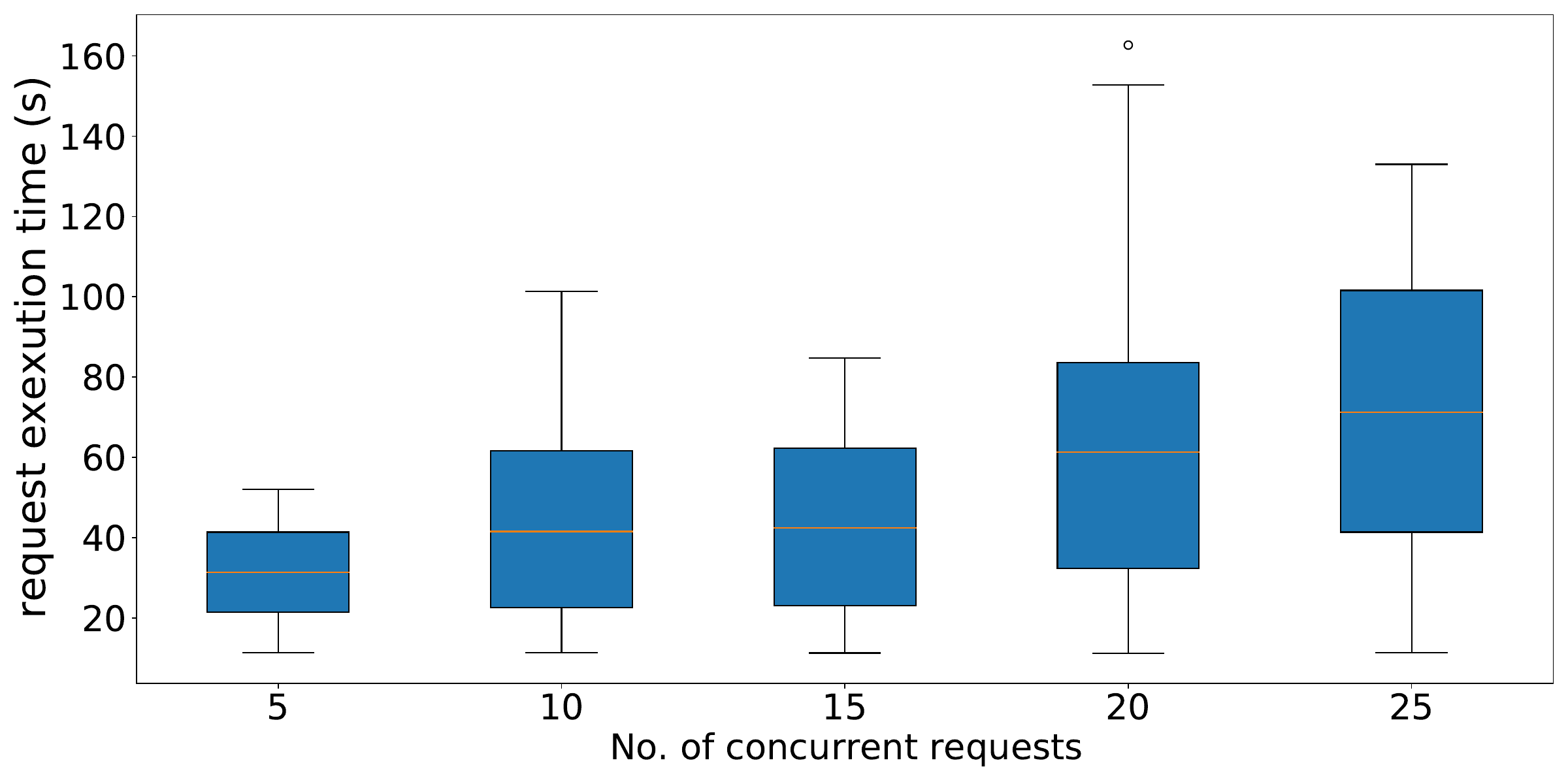}
		\end{minipage}
		\label{fig:azure-1024-performace}
}
\caption{The distribution of request execution time under different numbers of concurrent requests with 1,024 MB memory for four serverless computing platforms.}
\label{fig:concurrency1024performace}
\end{figure}

\begin{figure}[htbp]
\centering
\subfigure[AWS Lambda]{
		\begin{minipage}[b]{0.48\textwidth}
			\includegraphics[width=\textwidth]{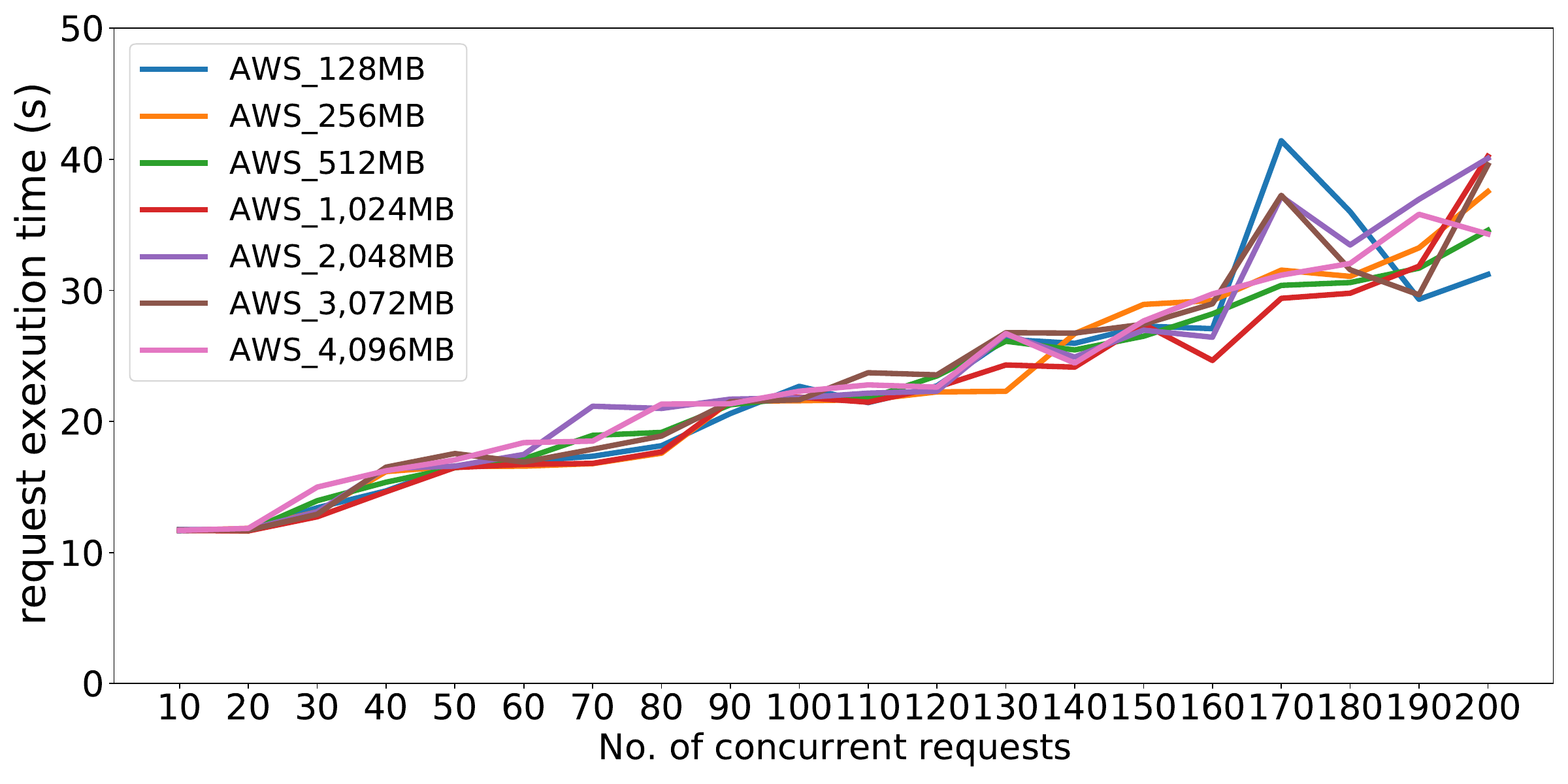}
		\end{minipage}
		\label{fig:aws-performace}
}
\subfigure[Google Cloud Functions]{
		\begin{minipage}[b]{0.48\textwidth}
			\includegraphics[width=\textwidth]{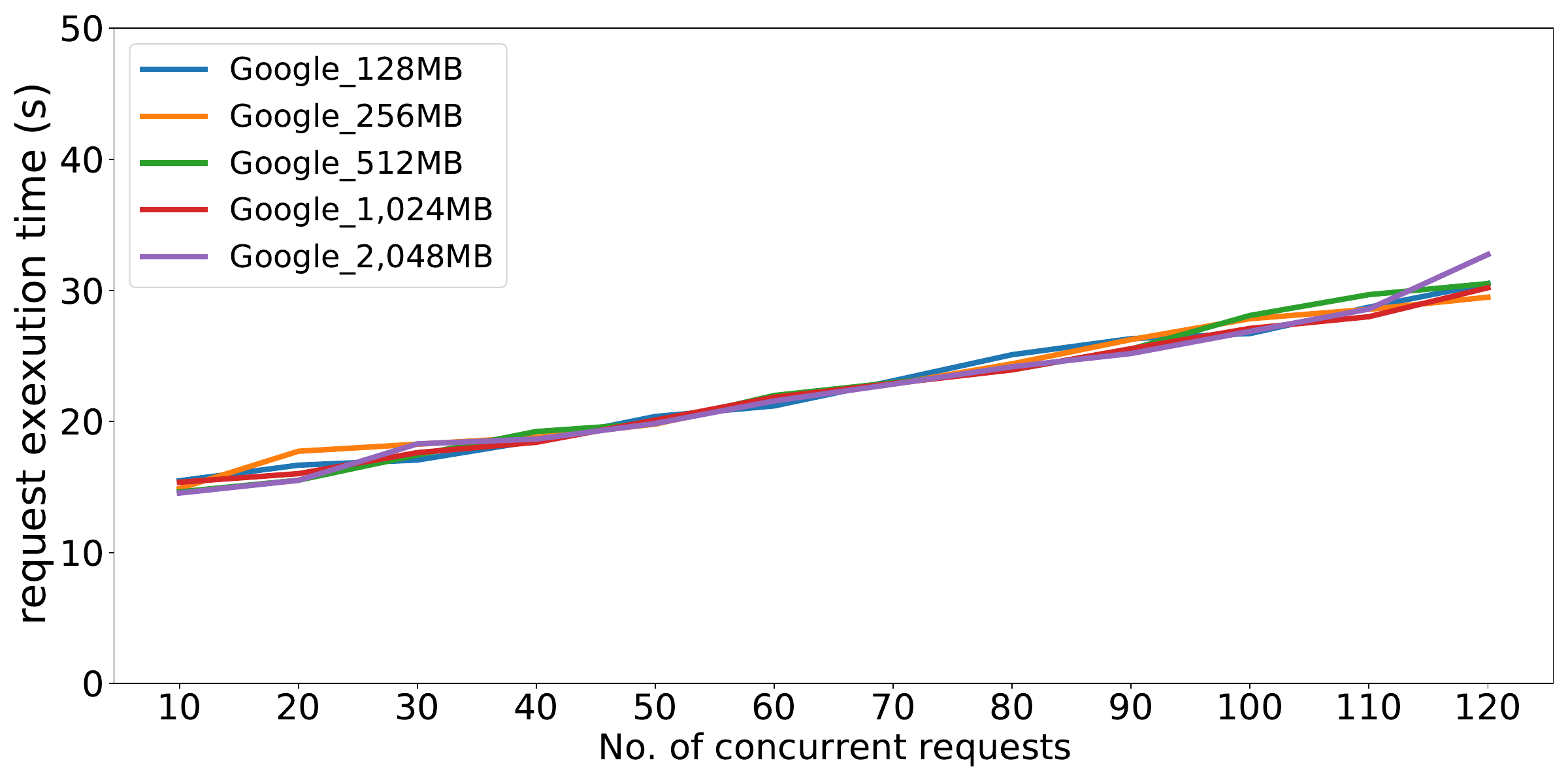}
		\end{minipage}
		\label{fig:google-performace}
}
\subfigure[Alibaba Cloud Function Compute]{
		\begin{minipage}[b]{0.48\textwidth}
			\includegraphics[width=\textwidth]{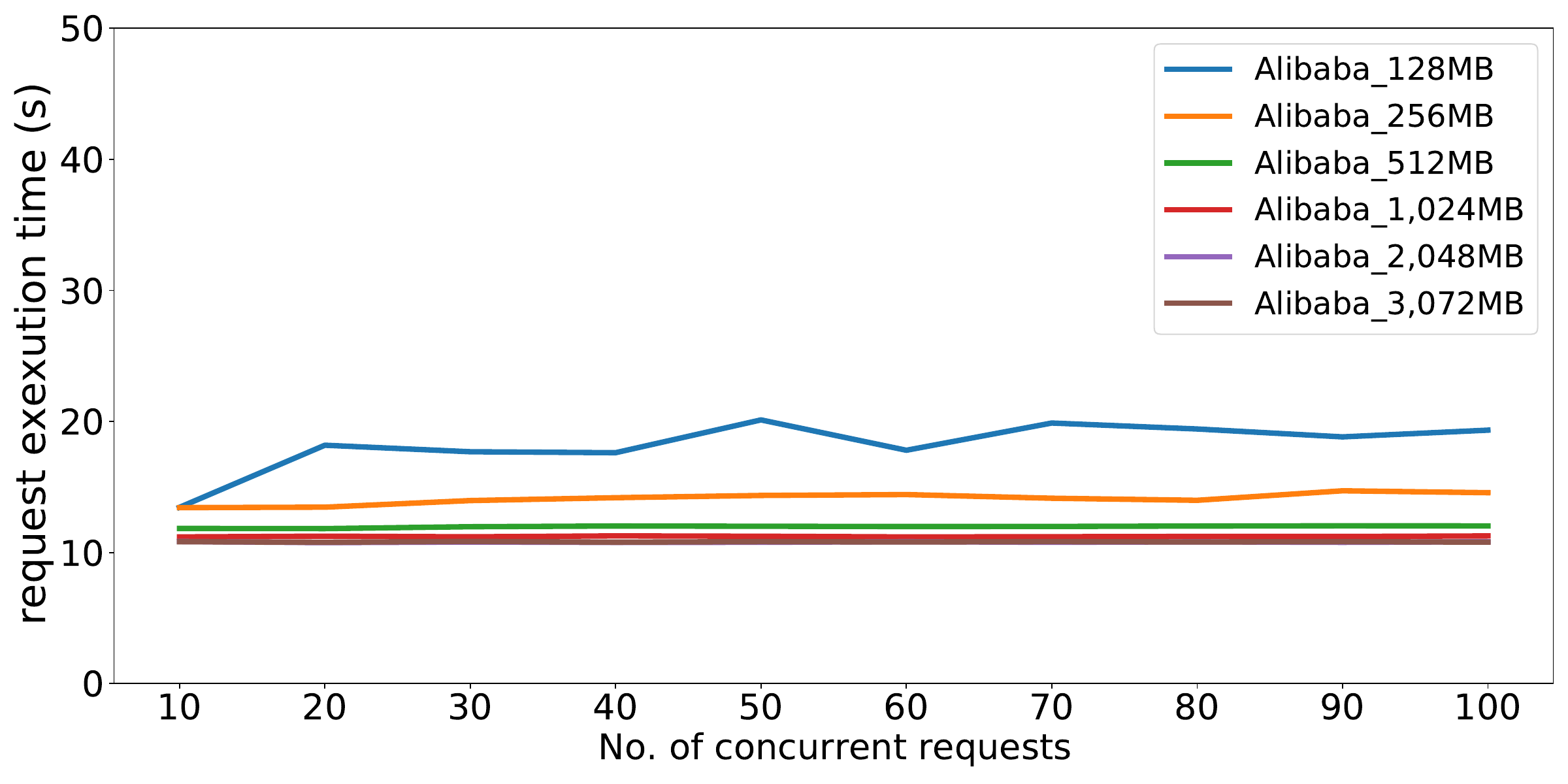}
		\end{minipage}
		\label{fig:ali-performace}
}
\subfigure[Azure Functions]{
		\begin{minipage}[b]{0.48\textwidth}
			\includegraphics[width=\textwidth]{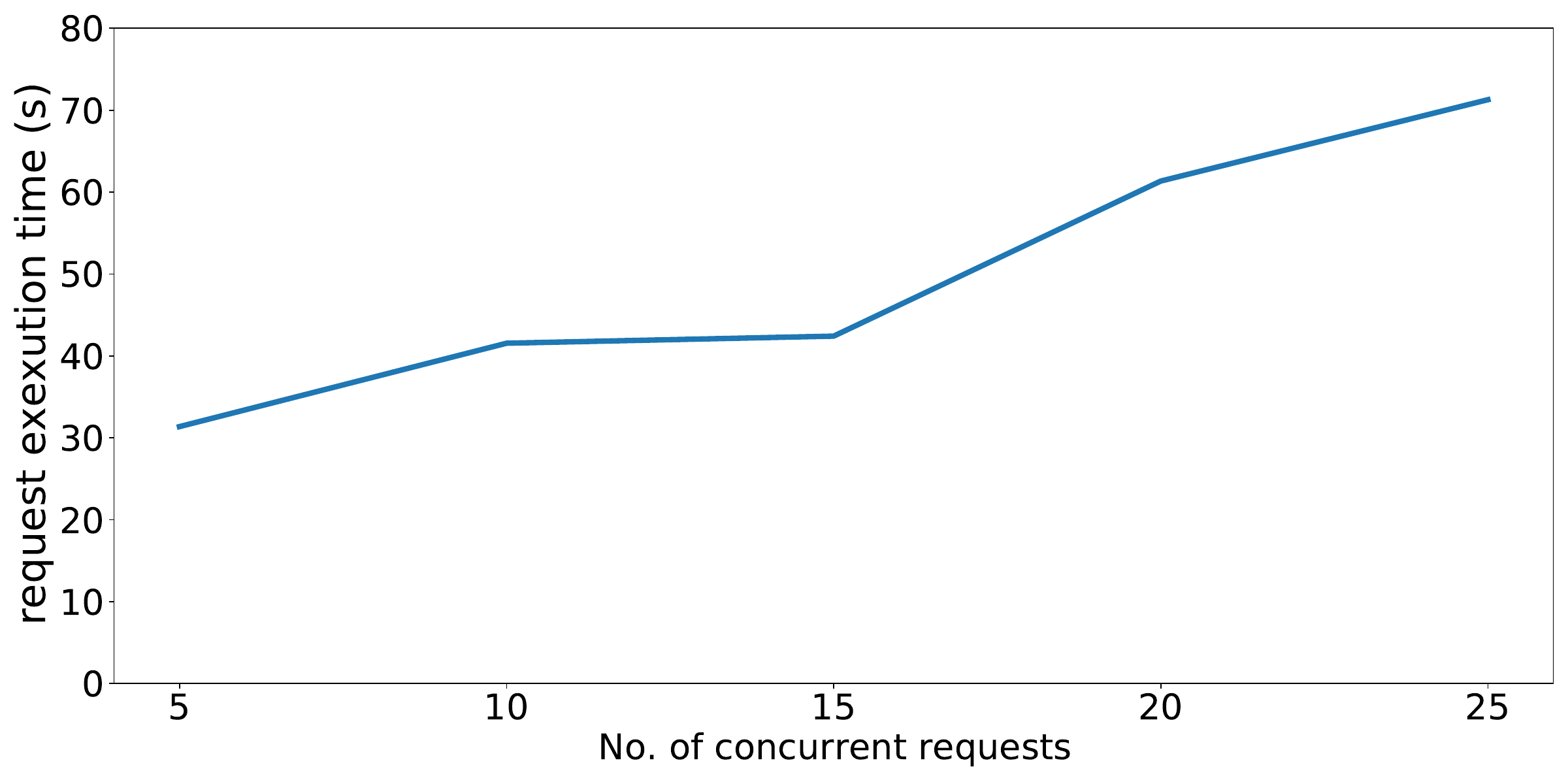}
		\end{minipage}
		\label{fig:azure-performace}
}
\caption{The median request execution time under different numbers of concurrent requests and various memory sizes for four serverless computing platforms.}
\label{fig:concurrencyperformace}
\end{figure}

In this section, we explore the concurrency performance of four serverless computing platforms. Meanwhile, we try to find the potential causes influencing concurrency performance for these platforms in a black-box fashion.

In concurrency experiments, we follow a similar approach in previous work\cite{WangATC2018, lloyd2018serverless} to create 20 python-based serverless functions with the same configuration and code but different function names [$f_{1}$, $f_{2}$, ..., $f_{20}$] and invoke each $f_{i}$ with 10$i$ concurrent requests. We pause for 20 seconds after each round of invocations to cope with rate limits in the platforms. All serverless functions execute our information collection function of \testbedName for further runtime analysis. Meanwhile, we also explore concurrency performance under various memory sizes. Particularly, Azure Functions cannot configure different memory sizes. The reason is that functions of Azure Functions are executed in the form of function applications, and there is no need to specify the memory size of the application in advance. In order to explore the concurrency performance of Azure Functions, we create and deploy functions with the same functionality as other platforms, without specifying the memory size. 


\subsubsection{Concurrency performance results}

Concurrency performance is represented as \emph{request execution time}, i.e., the time interval from the sending time of a request to the end time of processing this request. The results of concurrency performance are described as follows. Figure \ref{fig:concurrency1024performace} shows the distribution of request execution time under different numbers of concurrent requests with 1,204 MB memory for four serverless computing platforms. Figure \ref{fig:concurrencyperformace} shows the median request execution time under different numbers of concurrent requests and various memory sizes for four serverless computing platforms. From Figure \ref{fig:concurrency1024performace} and Figure \ref{fig:concurrencyperformace}, we conclude five experimental results. (1) AWS Lambda has the stronger ability to deal with a large number of bursty concurrent requests than other platforms. (2) The concurrency performance of Alibaba Cloud Function Compute does not change with different numbers of concurrent requests, and it also is the best and most stable among these platforms. Specifically, in Figure \ref{fig:concurrencyperformace}, compared with AWS Lambda, Google Cloud Functions, and Azure Functions, the concurrency performance of Alibaba Cloud Function Compute improves from 4.29\% to 102.50\%, from 37.27\% to 151.16\%, and from 271.31\% to 445.82\%, respectively. In addition, the variances of request execution time under different concurrency conditions on Alibaba Cloud Function Compute, AWS Lambda, Google Cloud Functions, and Azure Functions are 0.0007, 21.76, 52.48, and 211.38, respectively. The smaller the variance value means more stable the concurrency performance under different concurrency conditions. (3) AWS Lambda and Google Cloud Functions have similar concurrency performance in Figure \ref{fig:aws-performace} and Figure \ref{fig:google-performace}, and their request execution times increase linearly as the concurrency numbers increase. Moreover, AWS Lambda fluctuates in concurrency performance. We also find that the distribution of request execution time of AWS Lambda is more scattered by comparing the value of the ordinate in Figure \ref{fig:aws-1024-performace}. Note that in order to observe the trend of each platform at different concurrency conditions, we do not unify the ordinate values of Figure \ref{fig:concurrency1024performace}. (4) Azure Functions has the worst and most unstable concurrency performance among these platforms. For example, when there are 20 concurrent requests, the median request execution time of Azure Functions in Figure \ref{fig:azure-performace} is inferior to Alibaba Cloud Function Compute in Figure \ref{fig:ali-performace} by 445.82\%. (5) Different memory sizes affect the concurrency performance of Alibaba Cloud Function Compute, while other platforms are not affected. When allocated memory is smaller on Alibaba Cloud Function Compute, the concurrency performance is worse as shown in Figure \ref{fig:ali-performace}. For example, the performance of concurrent requests with 128 MB allocated memory is inferior to that with 3,072 MB allocated memory by about from 60\% to 80\%.

\noindent\textbf{Discussion and implications.} By comparing the concurrency performance under different concurrency conditions, we present some implications for developers. First, developers are advised to use AWS Lambda if they have a higher demand for the concurrency number. Second, developers who have requirements for high concurrency performance are advised to use Alibaba Cloud Function Compute. At the same time, they also need to consider the impact of memory on its concurrency performance. Using high memory can improve concurrency performance, but developers can have a trade-off between performance and cost. Additionally, if developers are using serverless computing platforms other than Alibaba Cloud Function Compute, they may not need to consider the impact of memory on concurrency performance. 


\subsubsection{Potential cause analysis}

We wonder about the potential causes that influence the concurrency performance of serverless computing platforms so that we can better explain the above results and findings. We explore the following three research questions, i.e., reasons that affect the number of concurrent tasks, the impact of scalability strategy on concurrency performance, and the impact of memory on concurrency performance of Alibaba Cloud Function Compute.

\noindent\textbf{$\bullet$ Reasons that affect the number of concurrent tasks.} In our experiments, we cannot conduct experiments under certain concurrency conditions for some platforms. In this section, we try to illustrate and explain them. Specifically, for Google Cloud Functions, when the number of concurrent requests is greater than 120, the invocation rate will exceed the pre-defined threshold and the platform will refuse to deal with incoming requests. Thus, we can only conduct experiments with concurrent requests increasing from 10 to 120on Google Cloud Functions. However, its official documentation claims that HTTP-triggered functions can scale to the desired invocation rate quickly\cite{googlescale}. There is an inconsistency between the official documentation and the actual runtime process. For Alibaba Cloud Function Compute, since the default number of concurrent requests supported by this platform is 100, we can only conduct experiments with concurrent requests increasing from 10 to 100. For Azure Functions, we find that it fails to execute concurrent requests greater than 25 under the consumption plan. The possible reason we think is that Azure Functions does not have the ability to provide the required resources for bursty workloads quickly. We have validated our conjecture (i.e., \emph{Azure Functions may not be able to handle bursty concurrency tasks well.}) through conducting the concurrency experiment with the ``large'' concurrency number (e.g., 200) in a ``warm'' state of Azure infrastructure. In this situation, we conduct only fine-grained bursty experiments for concurrent requests (e.g., increasing from 5 to 25) on Azure Functions to further explore its concurrency performance. 

\noindent\textbf{$\bullet$ The impact of scalability strategy on concurrency performance.} 
Generally, a function runs on dedicated instances like containers that provide the required runtime environment for this function. These instances are hosted on VMs. We try to analyze the runtime information, e.g., the usage of instances/VMs and the request distribution on VMs, to investigate the scalability strategy and load balancing ability of different platforms to further find potential causes influencing concurrency performance. The runtime information is collected by our collection function in \testbedName. The identification of VMs and instances, as well as the metric of requests distribution, are illustrated as follows.


\begin{itemize}
\item \emph{VMs identification.} For VMs identification, we leverage ``btime'' in \textbf{/proc/stat} to identify VMs as adopted in the previous work\cite{lloyd2018serverless}, which statistically demonstrates its high reliability.

\item \emph{Instances identification.} Following the previous work\cite{lloyd2018serverless, WangATC2018}, we check the existence of a temporary file on the local disk of the function instance when receiving a request. If the temporary file does not exist, we generate a universally unique 16-byte ASCII string to write to this file, which is served as the function instance ID. Since the local storage is non-persistent and has the same lifetime as the associated function instance, the temporary file will be removed once recycling the instance, but will not be modified or deleted when reusing existing instances.

\item \emph{Requests distribution.} We calculate the standard deviation of the numbers of requests distributed on different VMs to reflect the distribution of the requests. When the standard deviation of this distribution is zero, it implies that this platform will equably distribute incoming requests to multiple VMs. The increasing standard deviation implies the uneven distribution of requests on VMs. A single VM handling too many requests will endure resource competition and inefficiency, leading to performance degradation.
\end{itemize}

\begin{figure}
	\centering
	\subfigure[the number of the used function instances]{
		\begin{minipage}[b]{0.49\textwidth}
			\includegraphics[width=\textwidth]{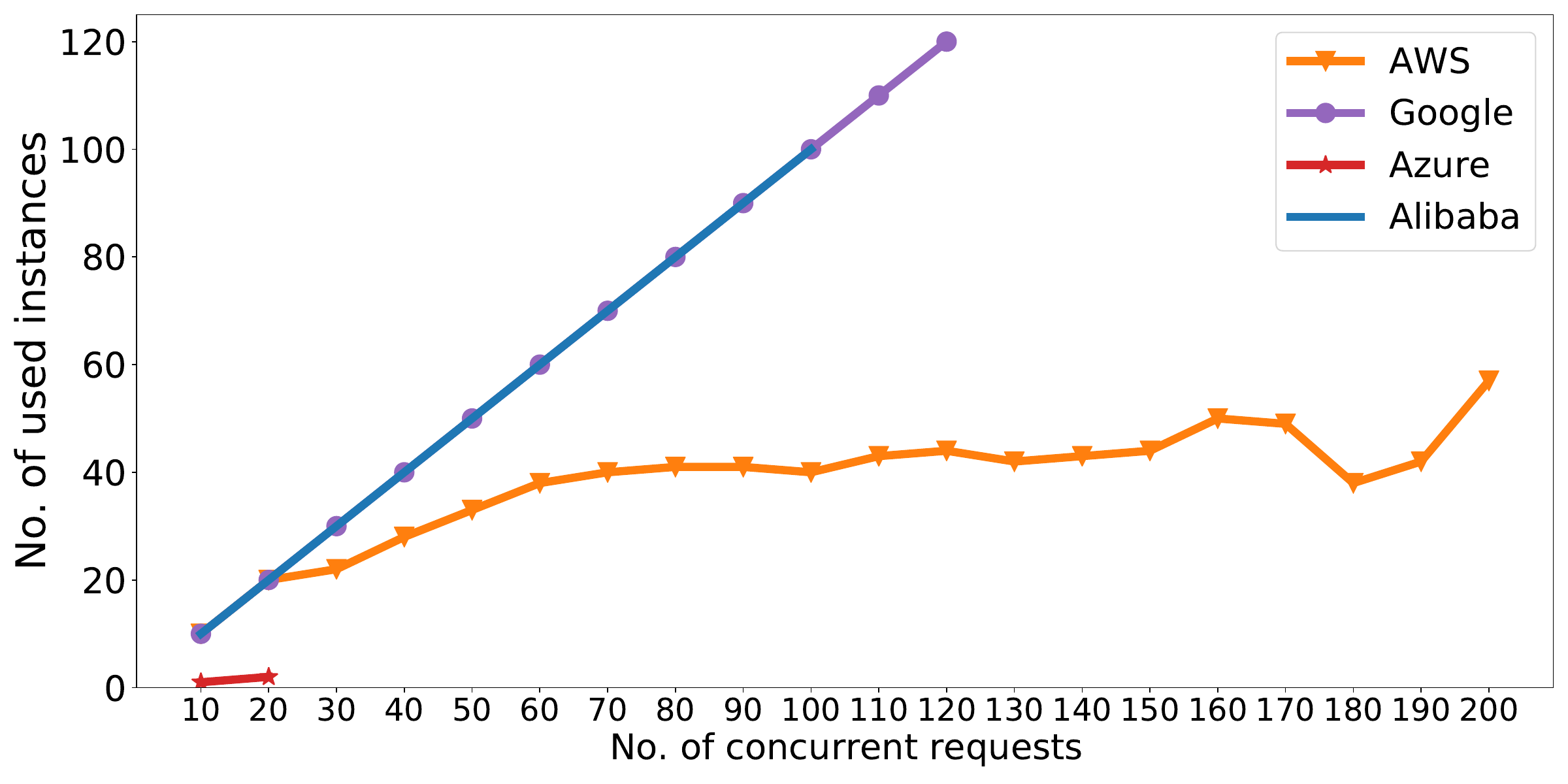}
			\label{fig:scalability_instance_No_1024MB}
		\end{minipage}
	}
    	\subfigure[the number of the used VMs]{
    		\begin{minipage}[b]{0.49\textwidth}
  		 	\includegraphics[width=\textwidth]{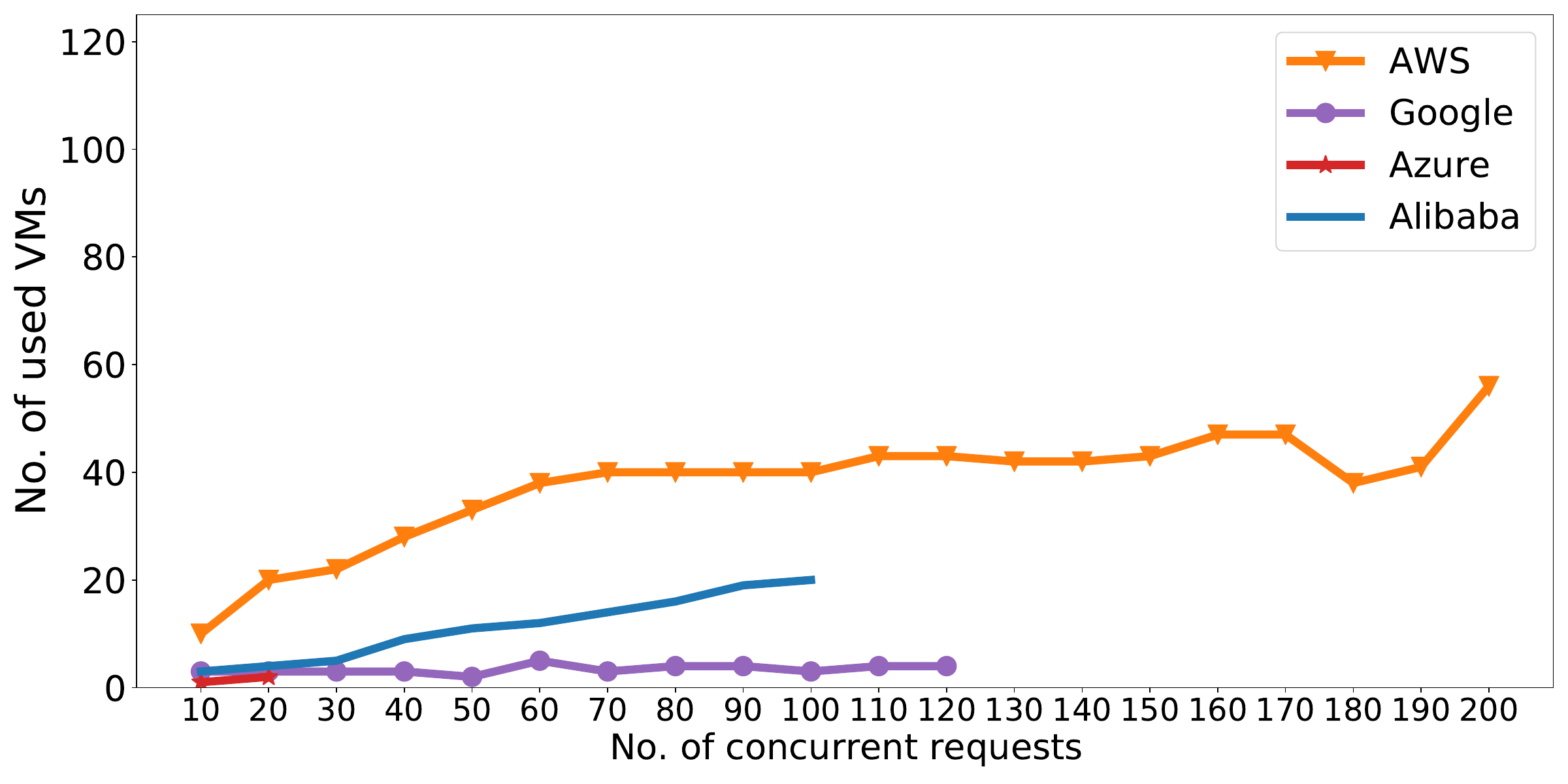}
  		 		\label{fig:scalability_VM_No_1024MB}
    		\end{minipage}
    	}
	\caption{The numbers of function instances and VMs used in different numbers of concurrent requests that each request allocates 1,024 MB memory.}
	\label{fig:scalability_No_1024MB}
\end{figure}

\begin{figure}
	\centering
	\subfigure[AWS]{
		\begin{minipage}[b]{0.3\textwidth}
			\includegraphics[width=1\textwidth]{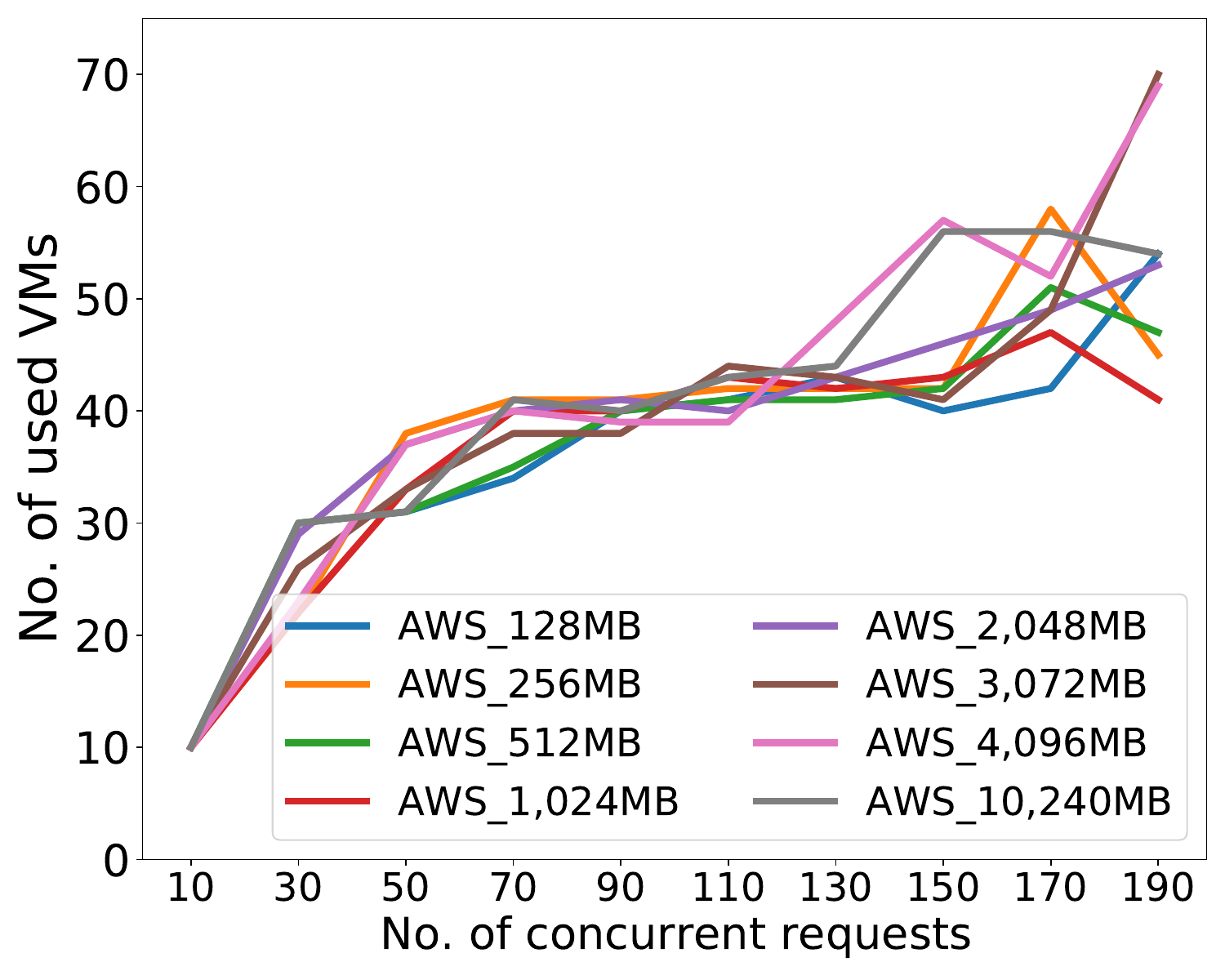}
		\end{minipage}
		\label{fig:scalability_memory_VMs_AWS}
	}
    	\subfigure[Google]{
    		\begin{minipage}[b]{0.3\textwidth}
  		 	\includegraphics[width=1\textwidth]{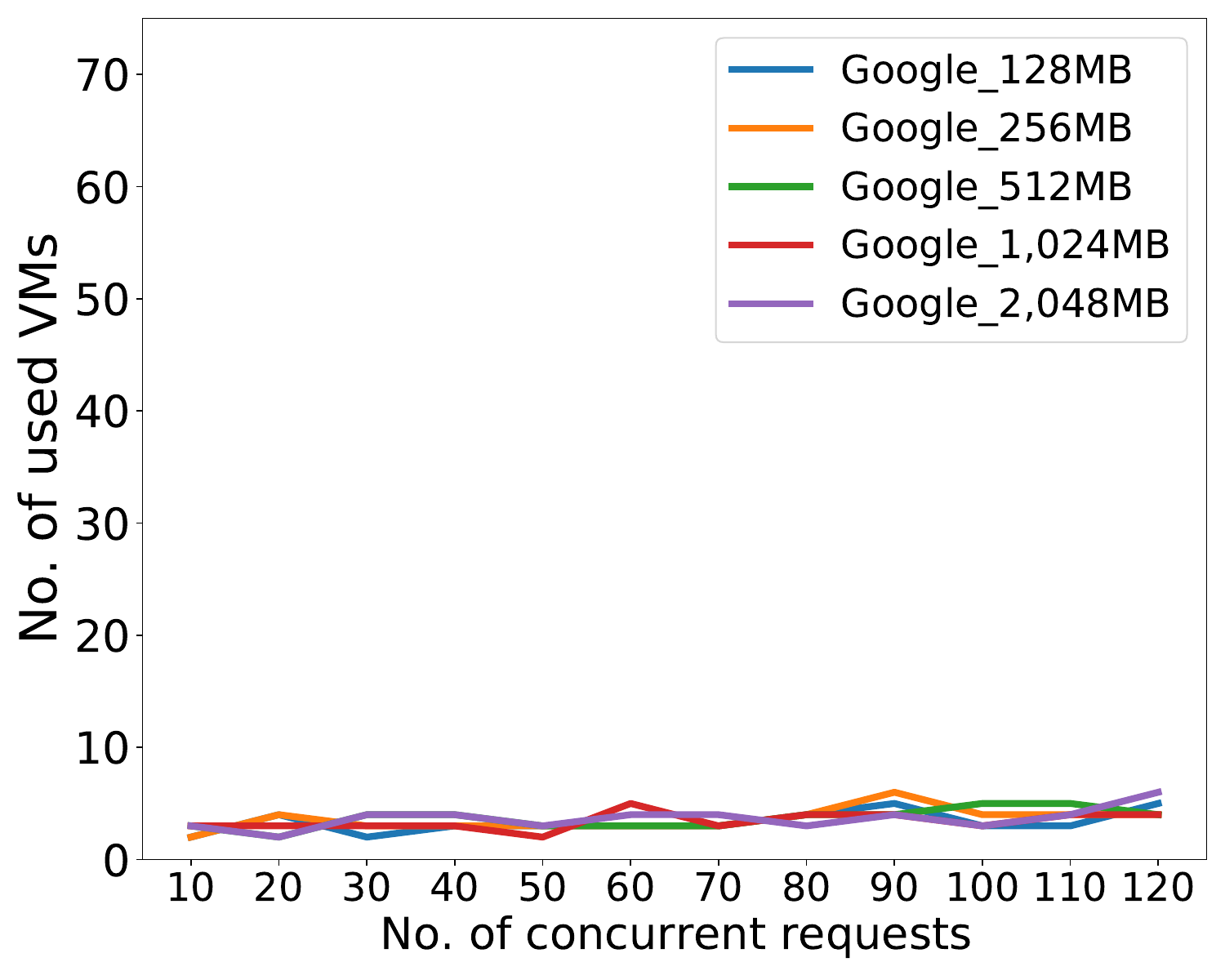}
    		\end{minipage}
		\label{fig:scalability_memory_VMs_Google}
    	}
    	\subfigure[Alibaba]{
    		\begin{minipage}[b]{0.3\textwidth}
  		 	\includegraphics[width=1\textwidth]{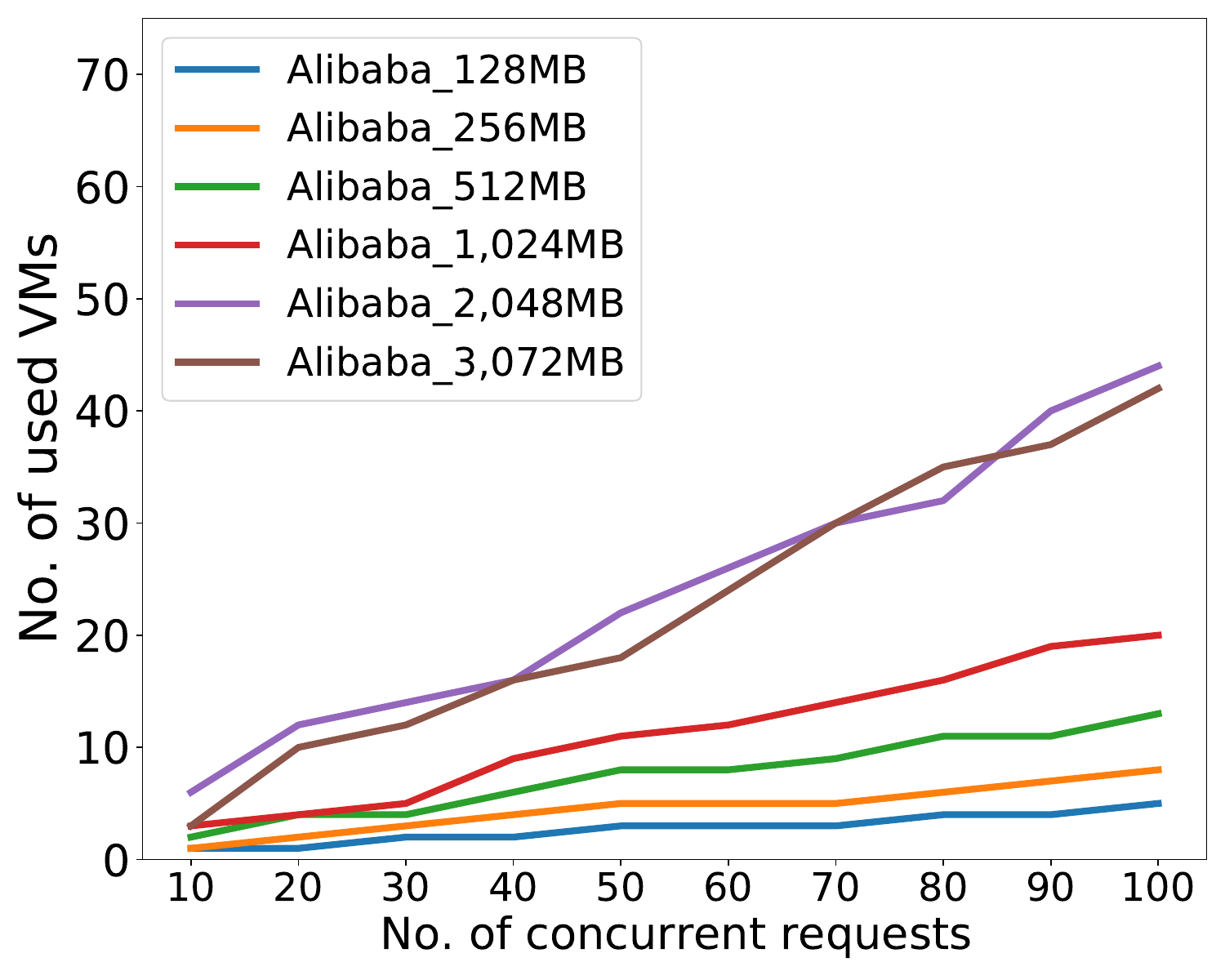}
    		\end{minipage}
		\label{fig:scalability_memory_VMs_Alibaba}
    	}
	\caption{The numbers of VMs used in different numbers of concurrent requests under various memory sizes.}
	\label{fig:scalability_memory_VMs}
\end{figure}

\begin{figure}
	\centering
	\subfigure[AWS]{
		\begin{minipage}[b]{0.3\textwidth}
			\includegraphics[width=1\textwidth]{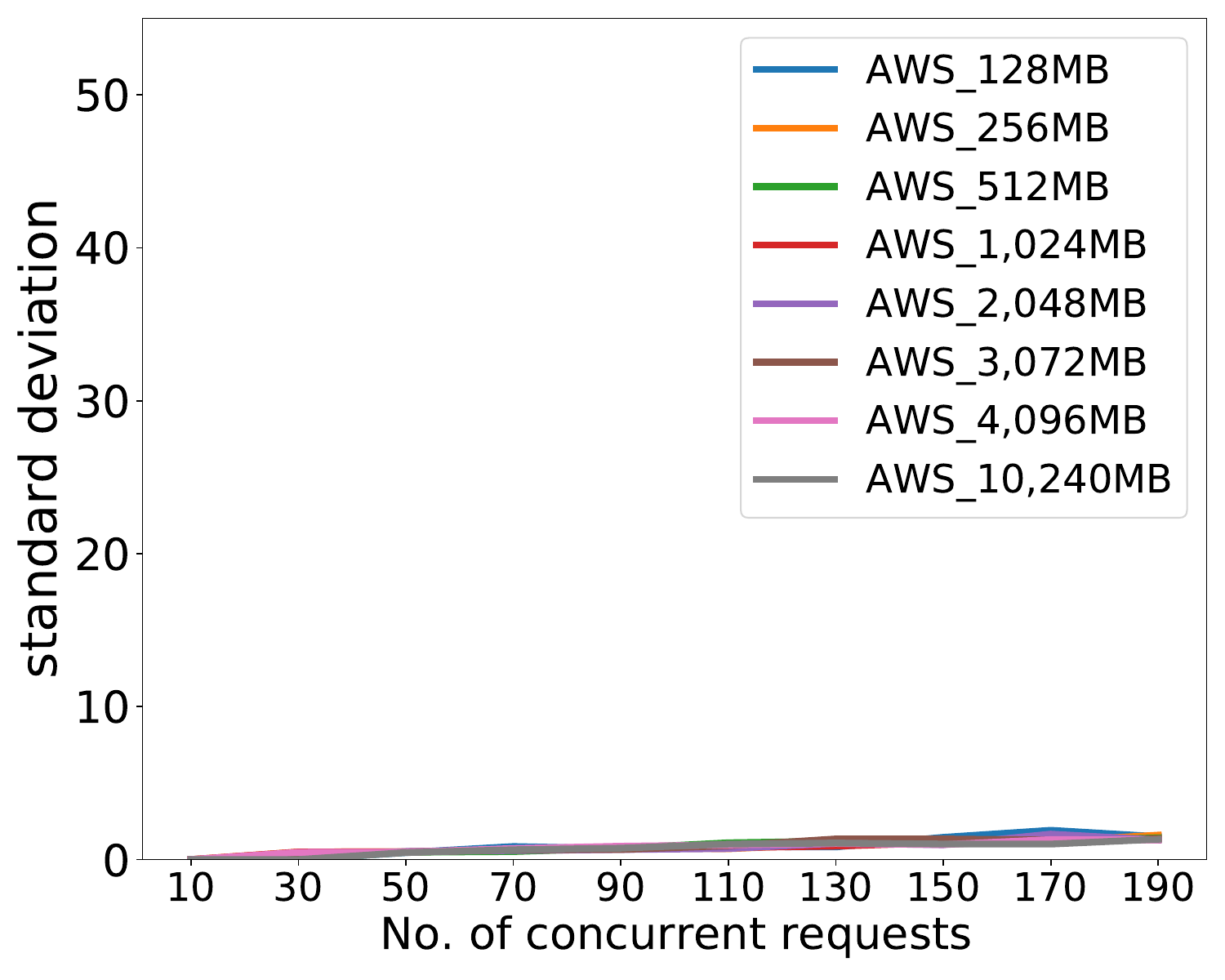}
		\end{minipage}
		\label{fig:scalability_memory_stand_dev_AWS}
	}
    	\subfigure[Google]{
    		\begin{minipage}[b]{0.3\textwidth}
  		 	\includegraphics[width=1\textwidth]{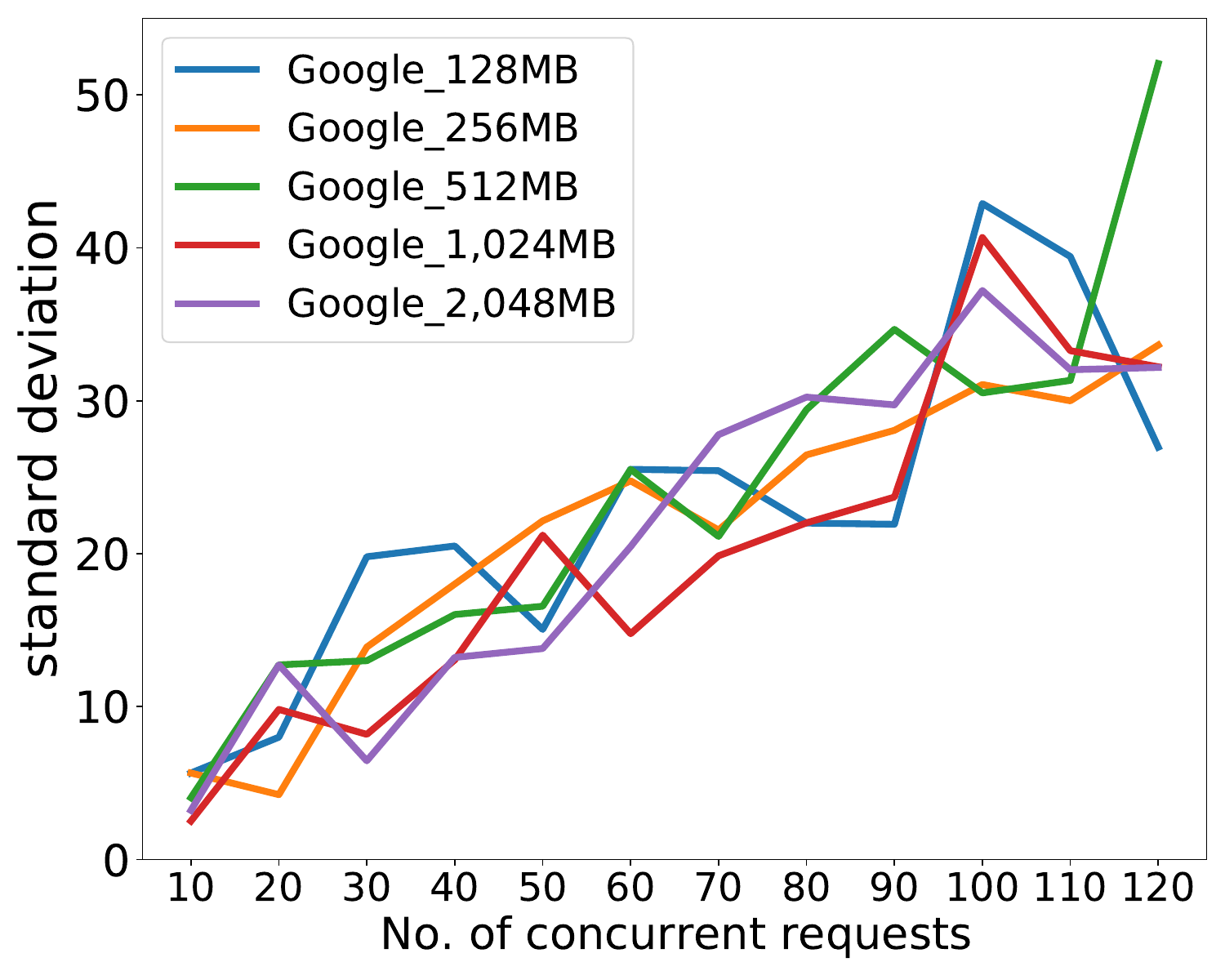}
    		\end{minipage}
		\label{fig:scalability_memory_stand_dev_Google}
    	}
    	\subfigure[Alibaba]{
    		\begin{minipage}[b]{0.3\textwidth}
  		 	\includegraphics[width=1\textwidth]{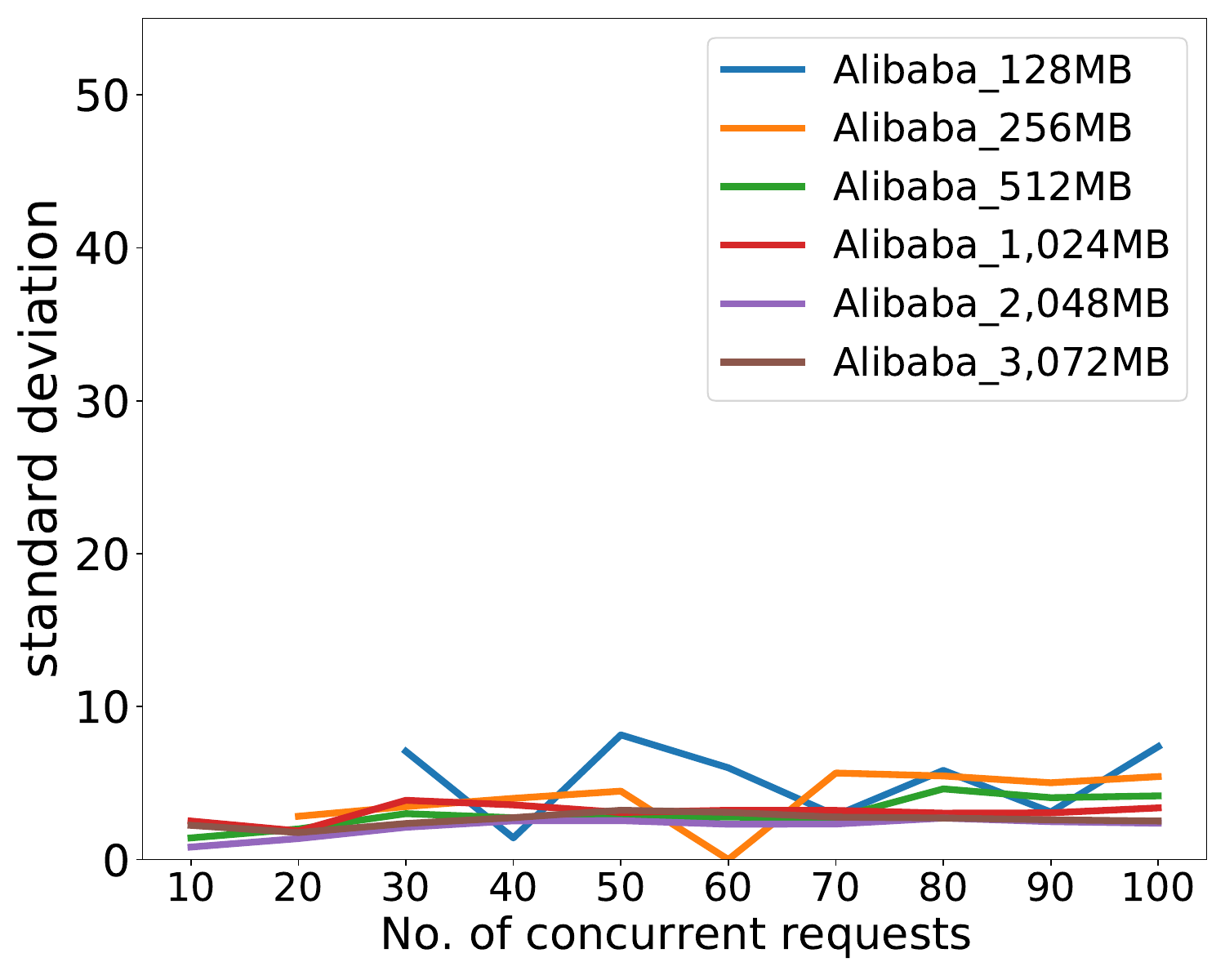}
    		\end{minipage}
		\label{fig:scalability_memory_stand_dev_Alibaba}
    	}
	\caption{The standard deviation of different numbers of concurrent requests under various memory sizes.}
	\label{fig:scalability_memory_stand_dev}
\end{figure}

We show the numbers of function instances and VMs under different concurrency conditions with 1,024 MB memory allocation (as shown in Figure \ref{fig:scalability_No_1024MB}), the numbers of VMs under different concurrency conditions with various memory sizes (shown in Figure \ref{fig:scalability_memory_VMs}), and the standard deviation under different concurrency conditions with various memory sizes (Figure \ref{fig:scalability_memory_stand_dev}).


The reason why Alibaba Cloud Function Compute gains a better concurrency performance may be attributed to its scalability strategy. Specifically, Figure \ref{fig:scalability_instance_No_1024MB} shows that Alibaba Cloud Function Compute will launch the same number of function instances as the number of concurrent requests. It indicates that each request is processed in a new dedicated function instance, and there are no re-usages of instances, resulting in relatively ideal parallel processing. The number of VMs used in Alibaba Cloud Function Compute increases as the number of concurrent requests increases as shown in Figure \ref{fig:scalability_VM_No_1024MB}. In the case of 100 concurrent requests, 20 VMs are used, which indicates that multiple function instances will be launched on a single VM. Through sorting and analyzing the start time of the function execution, we infer the scalability strategy of Alibaba Cloud Function Compute may be as follows. \textbf{On Alibaba Cloud Function Compute, it will launch a small part of VMs, and equably launch new function instances on these VMs alternately for all concurrent requests.} 

\begin{figure}[htbp]
\centering
\begin{minipage}[t]{0.48\textwidth}
\centering
\includegraphics[width=\textwidth]{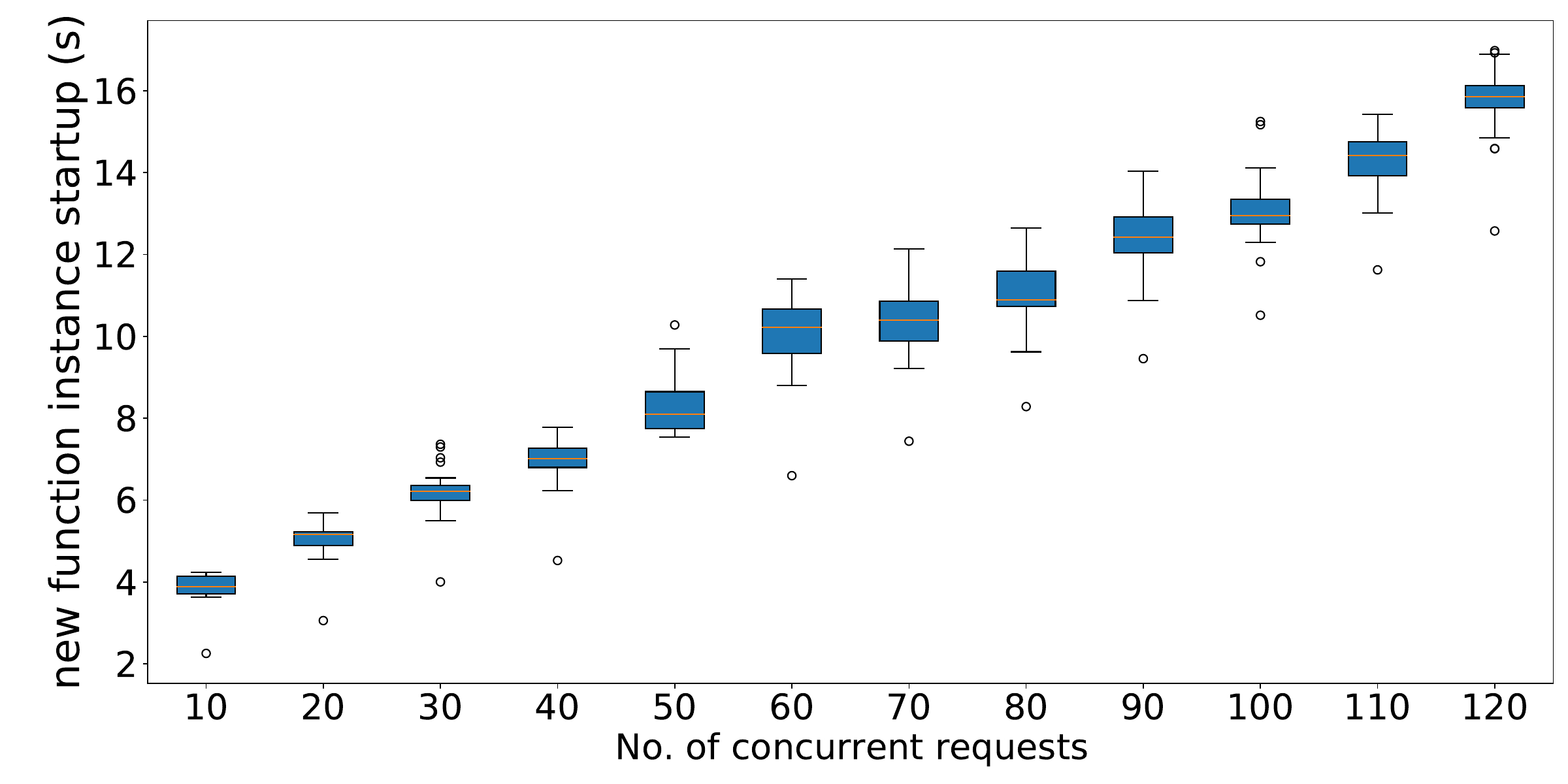}
\caption{The startup time of new function instances in different numbers of concurrent requests on Google Cloud Functions.}
\label{fig:scalability_startup_time_Google}
\end{minipage}
\begin{minipage}[t]{0.48\textwidth}
\centering
\includegraphics[width=\textwidth]{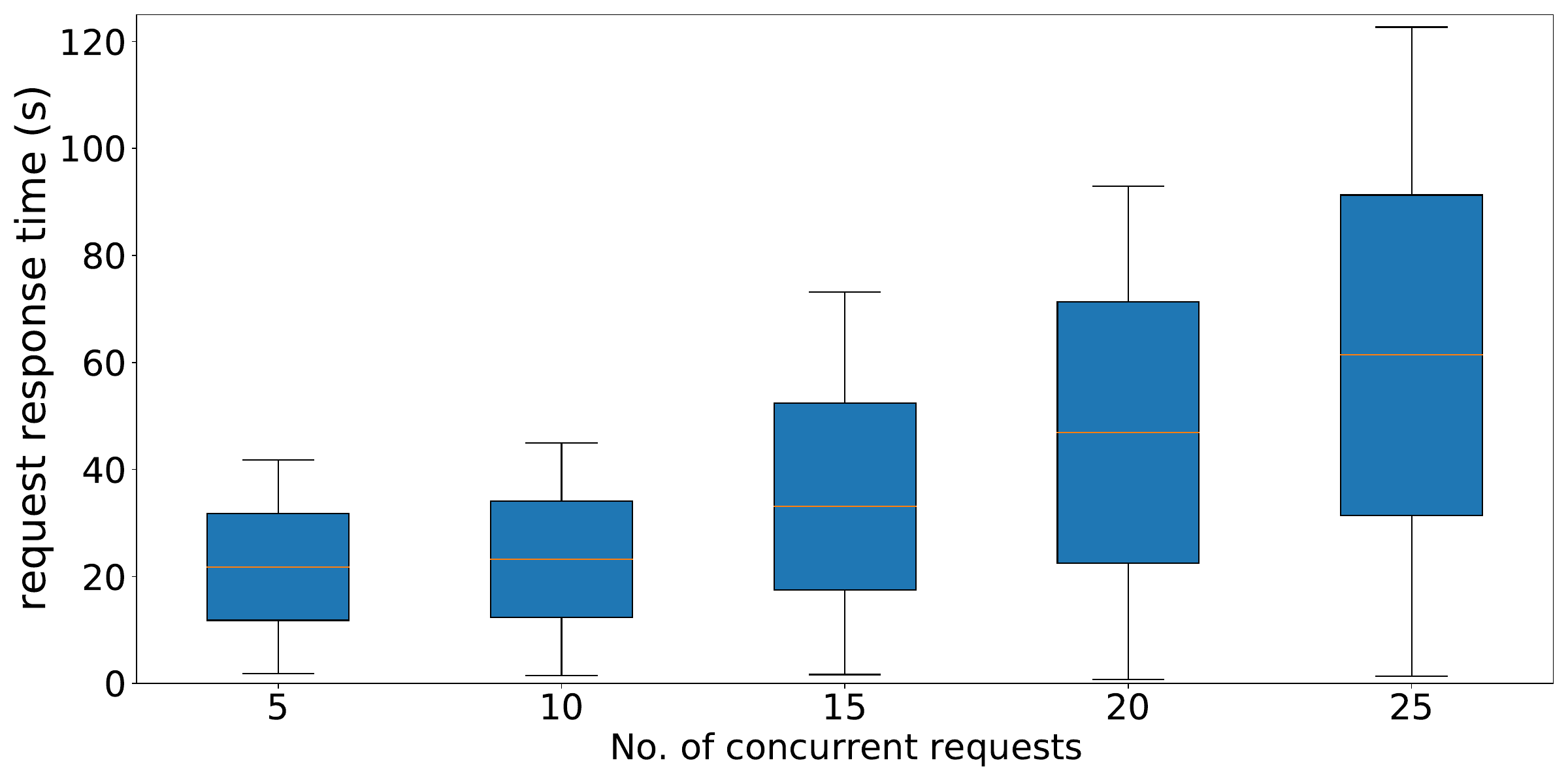}
\caption{The request response time in different numbers of concurrent requests on Azure Functions.}
\label{fig:scalability_request_time_Azure}
\end{minipage}
\end{figure}

Although AWS Lambda and Google Cloud Functions have similar concurrency performance, we find that they have different scalability strategies. For Google Cloud Functions, Figure \ref{fig:scalability_instance_No_1024MB} shows that its usage of function instances is the same as that of Alibaba Cloud Function Compute. However, the number of VMs used in Google Cloud Function is smaller and more stable than Alibaba Cloud Function Compute as shown in Figure \ref{fig:scalability_VM_No_1024MB}. It implies that more function instances are launched on a single VM on Google Cloud Functions. Through further analyzing the runtime information, we infer the scalability strategy of Google Cloud Functions may be as follows. \textbf{On Google Cloud Functions, it will generate a small number of VMs, and most new function instances will be launched on few VMs.} We explore the impact of this kind of strategy on the startup time of these new function instances, i.e., the time interval from the sending time of a request to the start processing time of its function instance. Figure \ref{fig:scalability_startup_time_Google} shows the startup time distribution of new function instances under different concurrency conditions with 1,024 MB memory. Overall, as the number of concurrent requests increases, the startup time of function instances will increases. It illustrates that the scalability strategy of Google Cloud Functions may affect the startup time of the function instance, and thus affect concurrency performance as shown in Figure \ref{fig:google-1024-performace}. Meanwhile, requests distribute more unevenly against the increase of concurrent requests as shown in Figure \ref{fig:scalability_memory_stand_dev_Google}. Thus, we conclude that \textbf{launching too many function instances in a single VM will lead to poor performance}.


For AWS Lambda, Figure \ref{fig:scalability_instance_No_1024MB} shows that the numbers of function instances and VMs are close under different concurrency conditions, which implies that only one function instance will be launched on a single VM. In addition, the number of function instances used by AWS Lambda is always less than the number of concurrent requests. For example, when the number of concurrent requests is 200, the number of function instances does not exceed 60. It indicates that AWS Lambda reuses existing function instances to deal with concurrent requests. Through further analyzing the runtime information, we infer the scalability strategy of AWS Lambda may be as follows. \textbf{On AWS Lambda, some requests will be executed on some function instances first, and other requests will wait for available instances. If there are available function instances, they will be reused to handle the remaining requests.} This kind of strategy can gain better load balancing as shown in Figure \ref{fig:scalability_memory_stand_dev_AWS}. In addition, the concurrency performance of AWS Lambda fluctuates with the increase of concurrent requests as in Figure \ref{fig:aws-performace}. The possible reason is the unstable of VMs startup as shown in Figure \ref{fig:scalability_memory_VMs_AWS} since their trends are similar.

For Azure Functions, through analyzing the results of concurrent requests less than or equal to 25, we find that the numbers of function instances and VMs used are the same, and at most two function instances or VMs are used. On one hand, the same numbers of function instances and VMs show that Azure Functions is similar to AWS Lambda, i.e., none of the concurrently running function instances are on the same VM. On the other hand, Azure Functions uses up to two function instances or VMs indicates that it cannot quickly launch enough function instances to handle all requests. In this situation, we infer that the scalability strategy may be as follows. \textbf{On Azure Functions, concurrent requests can be processed by only a small number of function instances at the same time, and the rest will wait.} Through observing the time interval at which the same function instance is used, the waiting time is about 10 seconds. We show the response time of requests (i.e., the time interval from the sending time of a request to the start time of the function contained in this request) under different concurrent numbers in Figure \ref{fig:scalability_request_time_Azure}. The response times of some requests increase as the increase of concurrency numbers. Thus, the reason that Azure Functions has a bad concurrency performance may also be attributed to its scalability strategy.

\noindent\textbf{$\bullet$ The impact of memory on concurrency performance of Alibaba Cloud Function Compute.}  In Figure \ref{fig:ali-performace}, when the allocated memory of concurrency tasks is smaller on Alibaba Cloud Function Compute, its concurrency performance is worse. Through observing the number of the used VMs in Figure \ref{fig:scalability_memory_VMs_Alibaba}, we find that this platform uses fewer VMs under the low memory condition than the high memory condition. Since the number of used function instances equals the number of concurrent requests under a certain concurrency task, this platform needs to launch more function instances on one VM in the low memory condition. However, this situation may increase the resource contention among function instances, resulting in performance degradation.

\noindent\textbf{Discussion and implications.} Through analyzing potential causes influencing concurrency performance, we present some practical implications for cloud vendors. First, cloud vendors are advised to improve their concurrency limits (e.g., the default concurrency number of Alibaba Cloud Function Compute), as well as solve the current shortcomings (e.g., the invocation ratio of Google Cloud Functions, and the ability to process bursty workloads on Azure Function). Second, these serverless computing platforms do not use function instances in the previous rounds except for Alibaba Cloud Function Compute. This kind of resource scheduling method may make the runtime environment of the function safer and prevent malicious attacks. However, if cloud vendors can solve security problems between function instances, they are advised to reuse function instances like Alibaba Cloud Function Compute to reduce the cold start and improve performance. Additionally, Google Cloud Functions and Alibaba Cloud Function Compute use a similar scalability strategy, but Google Cloud Functions launches most of the instances on few VMs. Due to resource competition among function instances, the performance will decrease. Thus, cloud vendors are advised to improve resource isolation among different instances.

\section{Discussion}\label{sec:discussion}

In this section, we discuss issues that may potentially affect the generalization of our results.

\noindent\textbf{$\bullet$ Other features of serverless computing platform.} In this paper, we do not investigate every aspect of serverless computing, such as permissions, security\cite{DattaWWW20}, monitoring, and so on. We mainly focus on features of configuration principle and actual runtime performance, in order to give practical implications for both developers and cloud vendors.

\noindent \textbf{$\bullet$ Fairness of evaluation.} In order to make a fair comparison, we deploy all functions in the same region (i.e., us-east in this paper) on all tested platforms. However, serverless computing platforms may have different performances in different regions. With \testbedName, researchers can conveniently repeat any experiment in other regions. We also plan to study how the deploying regions can affect the performance of applications.

\noindent\textbf{$\bullet$ Breakdown of the cold start time.}  The cold start may involve many processes, including downloading a package, launching a new container, setting up the runtime environment, and initializing the function. Although we have investigated many factors that may affect the overall cold start time, we fail to further locate the specific affected process due to the coarse-grained logs provided by these commodity serverless computing platforms. We plan to conduct a deeper analysis based on some open-source serverless computing platforms\cite{openWhisknew, fissionnew}.

\noindent\textbf{$\bullet$ Identification of VMs and function instances.} We explore the potential causes influencing concurrency performance via analyzing our collected runtime information. However, since the runtime systems of serverless computing platforms are not the same, we cannot directly obtain the accurate VM ID and function instance ID. In order to be able to compare their scalability strategies, we leverage the ``btime'' value of \textit{/proc/stat} or string of a temporary file to verify whether the used VMs or function instances are the same, respectively. These methods have also been adopted in the related work\cite{lloyd2018serverless, WangATC2018}.

\noindent\textbf{$\bullet$ The rapid evolution of serverless computing platforms.} The results and findings might change over time. These commodity serverless computing platforms are keeping adding new features and improving their performance, and later benchmarks on these serverless computing platforms may conflict with our current result. Fortunately, anyone can leverage \testbedName to benchmark target platforms continuously to keep tracing the up-to-date characteristics.



\section{Related Work}\label{sec:related_work}


Serverless computing is a new trending paradigm of cloud computing, and lots of related studies have been proposed.

Shahrad \textit{et al.}~\cite{ShahradATC20} characterized the entire production FaaS workload of Azure Functions, including trigger types, invocation frequencies and patterns, and resource needs. Spillner \textit{et al.}~\cite{SpillnerCoRR2019} conducted a quantitative study about how developers use FaaS offerings and gained insights into how functions are implemented and developed. However, these studies have not offered insights into the actual runtime performance that developers focus on. Meanwhile, these studies have not discussed the underlying features of scalability and load balancing in a black-box fashion, especially for commodity serverless computing platforms. In addition, some previous work~\cite{back2018using, figiela2018performance, lee2018evaluation, WangATC2018} has evaluated and compared how serverless applications behave on different platforms. Such work focused more on the workloads and the performance differences between platforms instead of qualitatively and quantitatively analyzing the underlying implications for developers and cloud vendors. Also, some work\cite{MohantyCloudCom18, PaladeService2019, LiWOSC2019} has focused on the performance of serverless computing with several popular open-source serverless computing platforms. Differently, we mainly focus on current mainstream commodity serverless computing platforms, and analyzing these platforms is more challenging because the underlying details are opaque to users.

In our paper, we also use some benchmarks to evaluate the actual runtime performance of serverless computing platforms. This point is motivated by the previous work. Yu \textit{et al.}~\cite{yu2020characterizing} proposed an open-source benchmark suite named \textit{ServerlessBench} to characterize serverless computing platforms leveraging customized test cases. Maissen \textit{et al.}~\cite{MaissenDEBSS2020} and Kim \textit{et al.}~\cite{KimLCLOUD19}  also designed two benchmark suites, named \textit{FaaSDOM} and \textit{FunctionBench}, respectively, to facilitate the performance testing of serverless computing platforms. They both provided microbenchmarks, and \textit{FunctionBench} provided some more complicated benchmarks that represent real-world applications. Differently, in \testbedName, we provide a more comprehensive benchmark suite than the previous work, enable developers or end-users to gain insights into the runtime actual performance of each platform. Currently, some work~\cite{eismann2020review, leitner2019mixed} has also analyzed the characteristics of serverless-based applications to guide the design of approaches related to serverless computing. In addition, Wen \textit{et al.}~\cite{WenServerless21} explored the specific challenges that developers encounter in developing serverless-based applications. Our work comprehensively explores the characteristics of serverless computing platforms from various aspects, i.e., basic static characteristics (in development, deployment, and runtime phases) and actual runtime performance (startup latency, resource efficiency, and concurrency performance). It can not only provide insightful guidance for serverless application development but also help cloud vendors to improve the related architecture design.





\section{Conclusion}\label{sec:conclusion}

In this paper, we have characterized four mainstream commodity serverless computing platforms (i.e., AWS Lambda, Google Cloud Functions, Azure Functions,  and Alibaba Cloud Function Compute) via qualitative analysis and quantitative analysis. Specifically, in qualitative analysis, we have compared serverless computing platforms from three aspects, i.e., development, deployment, and runtime, to construct a taxonomy of 20 characteristics. In quantitative analysis, we have developed an open-source evaluation tool \testbedName to explore the actual runtime performance of four serverless computing platforms from startup latency, execution latency, and scheduling latency. First, we have measured their startup latency from the key factors, e.g., programming languages, memory sizes, and package sizes. Second, we have characterized their resource efficiency using microbenchmarks and macrobenchmarks. Finally, we have explored their concurrency performance under different concurrent requests on these serverless computing platforms and further found the potential causes influencing their concurrency performance via analyzing their scalability and load balancing in a black-box fashion. Based on the results of both qualitative and quantitative analysis, we have provided a series of practical findings and actionable implications for developers and cloud vendors, intending to highlight good practices and interesting research avenues in adopting the serverless computing paradigm.




\section*{Acknowledgments}
This work was supported by the PKU-Baidu Fund Project under the grant number 2020BD007.


\bibliography{slsmeasure}

\begin{thebibliography}{10}
\providecommand \doibase [0]{http://dx.doi.org/}%

\bibitem{ao2018sprocket}
Ao L, Izhikevich L, Voelker GM, Porter G. Sprocket: A serverless video
  processing framework. In: Proceedings of the 2018 ACM Symposium on Cloud
  Computing. Association for Computing Machinery; 2018\string: 263--274.

\bibitem{carreira2019case}
Carreira J, Fonseca P, Tumanov A, Zhang A, Katz R. Cirrus: A serverless
  framework for end-to-end ML workflows. In: Proceedings of the ACM Symposium
  on Cloud Computing. Association for Computing Machinery; 2019\string: 13--24.

\bibitem{shankar2020numpywren}
Shankar V, Krauth K, Vodrahalli K, et al. Serverless linear algebra. In:
  Proceedings of the 2020 ACM Symposium on Cloud Computing. Association for
  Computing Machinery; 2020\string: 281--295.

\bibitem{gartner20new}
The CIO's guide to serverless computing.
  \url{https://www.gartner.com/smarterwithgartner/the-cios-guide-to-serverless-computing/};
  Retrieved on October 10, 2020.

\bibitem{JonasCoRR2019}
Jonas E, Schleier-Smith J, Sreekanti V, et al. Cloud programming simplified: A
  berkeley view on serverless computing. {\it arXiv preprint arXiv:1902.03383}
  2019.

\bibitem{AWSS3}
Amazon S3. \url{https://aws.amazon.com/cn/s3/};  Retrieved on October 10, 2020.

\bibitem{awsnew}
AWS Lambda. \url{https://docs.aws.amazon.com/lambda/latest/dg/welcome.html};
  Retrieved on October 10, 2020.

\bibitem{googlenew}
Google Cloud Functions. \url{https://cloud.google.com/functions};  Retrieved on
  October 10, 2020.

\bibitem{azurenew}
Azure Functions. \url{https://docs.microsoft.com/en-us/azure/azure-functions/};
   Retrieved on October 10, 2020.

\bibitem{alibabanew}
Alibaba Cloud Function Compute.
  \url{https://www.alibabacloud.com/products/function-compute};  Retrieved on
  October 10, 2020.

\bibitem{frey2013automatic}
Frey S, Hasselbring W, Schnoor B. Automatic conformance checking for migrating
  software systems to cloud infrastructures and platforms. {\it Journal of
  Software: Evolution and Process} 2013\string; 25(10)\string: 1089--1115.

\bibitem{Wen2021ServerlessWorkflow}
Wen J, Liu Y. A measurement study on serverless workflow services. In: 2021
  IEEE International Conference on Web Services (ICWS). IEEE. ; 2021\string:
  accepted to appear.

\bibitem{HellersteinCIDR19}
Hellerstein JM, Faleiro JM, Gonzalez J, et al. Serverless computing: one step
  forward, two steps back. In: Proceedings of the 9th Biennial Conference on
  Innovative Data Systems Research. www.cidrdb.org; 2019.

\bibitem{DuASPLOS2020}
Du D, Yu T, Xia Y, et al. Catalyzer: sub-millisecond startup for serverless
  computing with initialization-less booting. In: Proceedings of the 25th
  International Conference on Architectural Support for Programming Languages
  and Operating Systems. ACM; 2020\string: 467--481.

\bibitem{WenServerless21}
Wen J, Chen Z, Liu Y, et al. An empirical study on challenges of application
  development in serverless computing. In: Proceedings of the 28th {ACM} Joint
  Meeting on European Software Engineering Conference and Symposium on the
  Foundations of Software Engineering. ; 2021\string: 416--428.

\bibitem{lahmar2018multicloud}
Lahmar F, Mezni H. Multicloud service composition: a survey of current
  approaches and issues. {\it Journal of Software: Evolution and Process}
  2018\string; 30(10)\string: e1947.

\bibitem{toplanguagenew}
The PYPL popularity of programming language.
  \url{http://pypl.github.io/PYPL.html};  Retrieved on October 10, 2020.

\bibitem{tensorflownew}
Tensorflow 2.3.1-2.
  \url{https://www.archlinux.org/packages/community/x86_64/tensorflow/};
  Retrieved on October 10, 2020.

\bibitem{RomanoTSE20}
Romano S, Vendome C, Scanniello G, Poshyvanyk D. A multi-study investigation
  into dead code. {\it IEEE Transactions on Software Engineering} 2020\string;
  46(1)\string: 71--99.

\bibitem{Jun18PDPnew}
Kim J, Jun TJ, Kang D, Kim D, Kim D. GPU enabled serverless computing
  framework. In: Proceedings of the 26th Euromicro International Conference on
  Parallel, Distributed and Network-based Processing. IEEE; 2018\string:
  533--540.

\bibitem{alibabaGPUnew}
Alibaba elastic GPU service. \url{https://www.alibabacloud.com/product/gpu};
  Retrieved on October 10, 2020.

\bibitem{AgacheNSDI2020}
Agache A, Brooker M, Iordache A, et al. Firecracker: lightweight virtualization
  for serverless applications. In: Proceedings of the 17th USENIX Symposium on
  Networked Systems Design and Implementation. USENIX Association; 2020\string:
  419--434.

\bibitem{gvisornew}
gVisor is an application kernel for containers that provides efficient
  defense-in-depth anywhere. \url{https://gvisor.dev/};  Retrieved on October
  10, 2020.

\bibitem{fibonaccinew}
Fibonacci number. \url{https://en.wikipedia.org/wiki/Fibonacci_number};
  Retrieved on October 10, 2020.

\bibitem{Pillow}
Pillow. \url{https://pillow.readthedocs.io/en/stable/};  Retrieved on October
  10, 2020.

\bibitem{OpenCV}
OpenCV. \url{https://opencv.org/};  Retrieved on October 10, 2020.

\bibitem{SqueezeNet}
SqueezeNet. \url{https://en.wikipedia.org/wiki/SqueezeNet};  Retrieved on
  October 10, 2020.

\bibitem{ImageNet}
ImageNet. \url{https://en.wikipedia.org/wiki/ImageNet};  Retrieved on October
  10, 2020.

\bibitem{TensorflowKeras}
Tensorflow Keras. \url{https://www.tensorflow.org/api_docs/python/tf/keras?en};
   Retrieved on October 10, 2020.

\bibitem{CNN}
Convolutional neural network.
  \url{https://en.wikipedia.org/wiki/Convolutional_neural_network};  Retrieved
  on October 10, 2020.

\bibitem{RNN}
Recurrent neural network.
  \url{https://en.wikipedia.org/wiki/Recurrent_neural_network};  Retrieved on
  October 10, 2020.

\bibitem{PyTorch}
PyTorch. \url{https://pytorch.org/};  Retrieved on October 10, 2020.

\bibitem{OakesATC18}
Oakes E, Yang L, Zhou D, et al. SOCK: rapid task provisioning with
  serverless-optimized containers. In: Proceedings of the 2018 USENIX Annual
  Technical Conference. USENIX Association; 2018\string: 57--70.

\bibitem{WangATC2018}
Wang L, Li M, Zhang Y, Ristenpart T, Swift M. Peeking behind the curtains of
  serverless platforms. In: Proceedings of the 2018 USENIX Annual Technical
  Conference. ; 2018\string: 133--146.

\bibitem{iPerf3}
iPerf3. \url{https://iperf.fr/iperf-download.php};  Retrieved on October 10,
  2020.

\bibitem{lloyd2018serverless}
Lloyd W, Ramesh S, Chinthalapati S, Ly L, Pallickara S. Serverless computing:
  An investigation of factors influencing microservice performance. In:
  Proceedings of 2018 IEEE International Conference on Cloud Engineering. IEEE;
  2018\string: 159--169.

\bibitem{googlescale}
The quota limits for Google Cloud Functions.
  \url{https://cloud.google.com/functions/quotas};  Retrieved on October 10,
  2020.

\bibitem{DattaWWW20}
Datta P, Kumar P, Morris T, Grace M, Rahmati A, Bates A. Valve: securing
  function workflows on serverless computing platforms. In: Proceedings of the
  29th International Conference on World Wide Web. Association for Computing
  Machinery; 2020\string: 939--950.

\bibitem{openWhisknew}
Apache openWhisk is a serverless, open source cloud platform that allows you to
  execute code in response to events at any scale.
  \url{https://openwhisk.apache.org/};  Retrieved on October 10, 2020.

\bibitem{fissionnew}
Fission: Serverless Functions for Kubernetes.
  \url{https://github.com/fission/fission};  Retrieved on October 10, 2020.

\bibitem{ShahradATC20}
Shahrad M, Fonseca R, Goiri {\'I}, et al. Serverless in the wild:
  characterizing and optimizing the serverless workload at a large cloud
  provider. In: Proceedings of the 2020 USENIX Annual Technical Conference.
  {USENIX} Association; 2020\string: 205--218.

\bibitem{SpillnerCoRR2019}
Spillner J. Quantitative snalysis of cloud function evolution in the AWS
  serverless application repository. {\it CoRR} 2019\string; abs/1905.04800.

\bibitem{back2018using}
Back T, Andrikopoulos V. Using a microbenchmark to compare function as a
  service solutions. In: Proceedings of the European Conference on
  Service-Oriented and Cloud Computing. Springer; 2018\string: 146--160.

\bibitem{figiela2018performance}
Figiela K, Gajek A, Zima A, Obrok B, Malawski M. Performance evaluation of
  heterogeneous cloud functions. {\it Concurrency and Computation: Practice and
  Experience} 2018\string; 30(23)\string: e4792.

\bibitem{lee2018evaluation}
Lee H, Satyam K, Fox GC. Evaluation of production serverless computing
  environments. In: Proceedings of the 2019 IEEE International Conference on
  Cloud Computing. IEEE Computer Society; 2018\string: 442--450.

\bibitem{MohantyCloudCom18}
Mohanty SK, Premsankar G, Francesco MD. An evaluation of open source serverless
  computing frameworks. In: Proceedings of the 2018 IEEE International
  Conference on Cloud Computing Technology and Science. IEEE Computer Society;
  2018\string: 115--120.

\bibitem{PaladeService2019}
Palade A, Kazmi A, Clarke S. An evaluation of open source serverless computing
  frameworks support at the edge. In: Proceedings of the 2019 IEEE World
  Congress on Services. {IEEE}; 2019\string: 206--211.

\bibitem{LiWOSC2019}
Li J, Kulkarni SG, Ramakrishnan KK, Li D. Understanding open source serverless
  platforms: design considerations and performance. In: Proceedings of the 5th
  International Workshop on Serverless Computing. ACM; 2019\string: 37--42.

\bibitem{yu2020characterizing}
Yu T, Liu Q, Du D, et al. Characterizing serverless platforms with
  serverlessbench. In: Proceedings of the 11th ACM Symposium on Cloud
  Computing. Association for Computing Machinery; 2020\string: 30--44.

\bibitem{MaissenDEBSS2020}
Maissen P, Felber P, Kropf P, Schiavoni V. FaaSdom: A benchmark suite for
  serverless computing. In: Proceedings of the 14th {ACM} International
  Conference on Distributed and Event-based Systems. {ACM}; 2020\string:
  73--84.

\bibitem{KimLCLOUD19}
Kim J, Lee K. Functionbench: A suite of workloads for serverless cloud function
  service. In: Proceedings of the 2019 IEEE International Conference on Cloud
  Computing. {IEEE}; 2019\string: 502--504.

\bibitem{eismann2020review}
Eismann S, Scheuner J, Eyk vE, et al. A review of serverless use cases and
  their characteristics. {\it arXiv preprint arXiv:2008.11110} 2020.

\bibitem{leitner2019mixed}
Leitner P, Wittern E, Spillner J, Hummer W. A mixed-method empirical study of
  Function-as-a-Service software development in industrial practice. {\it
  Journal of Systems and Software} 2019\string; 149\string: 340--359.

\end{thebibliography}



\end{document}